\documentclass[12pt,preprint]{aastex}

\usepackage{multirow}

\shorttitle{ALMA Observations of Ethyl Formate toward Orion KL}

\shortauthors{Peng et al.}

\begin{document}




\title{ALMA Observations of Ethyl Formate toward Orion KL}







\author{Y\sc{aping} P\sc{eng}\altaffilmark{1,2}, V. M. R\sc{ivilla}\altaffilmark{3}, L\sc{i} Z\sc{hang}\altaffilmark{1}, J. X. G\sc{e}\altaffilmark{4}, B\sc{ing} Z\sc{hou}\altaffilmark{1,2} }
\altaffiltext{1}{Department of
Astronomy, Yunnan University, and  Key Laboratory of Astroparticle Physics of
Yunnan Province, Kunming, 650091, China; pyp893@163.com, lizhang@ynu.edu.cn, bingzhoukm@126.com}
\altaffiltext{2}{College of Science, Yunnan Agricultural University, Kunming, 650201, China}
\altaffiltext{3}{INAF-Osservatorio Astrofisico di Arcetri, Largo Enrico Fermi 5, I-50125, Firenze, Italia; rivilla@arcetri.astro.it}
\altaffiltext{4}{Departamento de Astronom\'{\i}a, Universidad de Chile, Camino el Observatorio 1515, Las Condes, Santiago, Chile; gejixing666@gmail.com}






\begin{abstract}
Orion KL is one of the prime templates of astrochemical and prebiotic chemical studies. We wish to explore more organic molecules with increasing complexity in this region. In particular, we have searched for one of the most complex organic molecules detected in space so far, ethyl formate (C$_{2}$H$_{5}$OCHO). This species is the next step in chemical complexity after the simplest member of esters (methyl formate, CH$_{3}$OCHO). The mechanisms leading to its formation are still poorly known. We have used high angular resolution ($\sim$ 1.$^{\prime\prime}$5) ALMA observations covering a large bandwidth from 214 to 247 GHz. We have detected 82 unblended lines of C$_{2}$H$_{5}$OCHO (49 and 33 of the trans and gauche conformers, respectively). The line images showed that C$_{2}$H$_{5}$OCHO arises mainly from the compact ridge and the hot core-southwest regions. The derived rotational temperatures and column densities are 122 $\pm$ 34 K, (0.9 $\pm$ 0.3) $\times$ 10$^{16}$ cm$^{-2}$ for the hot core-SW, and 103 $\pm$ 13 K, (0.6 $\pm$ 0.3) $\times$ 10$^{16}$ cm$^{-2}$ for the compact ridge. The comparison of spatial distribution and abundance ratios with chemically related molecules (methyl formate, ethanol and formic acid) indicates that C$_{2}$H$_{5}$OCHO is likely formed on the surface of dust grains by addition of CH$_{3}$ to functional-group radicals (CH$_{2}$OCHO) derived from methyl formate (CH$_{3}$OCHO).

\end{abstract}














\section{INTRODUCTION}\label{s:intro}
 Almost 200 molecules have been detected in the interstellar medium (ISM) and circumstellar environments, of which more than 65 are considered complex organic molecules (COMs), molecules with $\geq$ 6 atoms containing carbon. The formation of these COMs in the ISM is still poorly understood, and is subject of intense debate. Two general theories have been proposed: gas-phase chemistry (e.g., Millar et al. 1991; Balucani et al. 2015; Taquet et al. 2016) and dust grain surface chemistry (e.g., Garrod \& Herbst 2006; Fedoseev et al. 2015; Chuang et al. 2016).

 Orion KL is a high mass star-forming region that provides a superb template for constraining the chemical models (e.g., \citealt{Brouillet13,Crockett15,Feng15,Favre17,Carroll17,Friedel17}). Orion KL consists of different spatial and chemical components, namely: the hot core, the compact ridge, the extended ridge and the plateau \citep{Blake87}. In particular two regions, the hot core and the compact ridge, have received special attention because they are very rich in organic species (e.g., Blake et al.\ 1987; Wang et al.\ 2011; Favre et al.\ 2011). While the hot core is rich in nitrogen-bearing molecules \citep{Peng13,Peng17,Pagani17}, the compact ridge is richer in oxygen-bearing species \citep{Blake87,Beuther05,Friedel08,Gong15}. The hot core region may be heated by shocks from source I outflow (e.g., Goddi et al.\ 2011) or by explosive event occurred around 500 yr ago (e.g., Zapata et al.\ 2011; Rivilla et al.\ 2013; Bally et al.\ 2011, 2017; Wright \& Plambeck 2017). Whether this explosive event impacted the release of complex molecules to the gas phase is under debate (e.g., Friedel \& Looney 2017; Favre et al.\ 2017).

The ester family of molecules is essential for many prebiotic processes. The simplest representative is the 8 atom molecule methyl formate (CH$_{3}$OCHO, hereafter MF), which is relatively abundant in many high-mass (e.g., Favre et al. 2011; Rivilla et al. 2017b) and low-mass star-forming regions (e.g., Cazaux et al. 2003). Ethyl formate (C$_{2}$H$_{5}$OCHO, hereafter EF) is the next step in chemical complexity of esters, with eleven atoms. EF has two conformers: the $trans$-conformer (also called $anti$-conformer) where the heavy atoms C-C-O-C=O form a planar zigzag chain, and the $gauche$-conformer produced by an out-of-plane rotation of the ethyl group (C$_{2}$H$_{5}$) around the O-C bond \citep{Medvedev09}. The formation routes of EF are still poorly known. Several chemical pathways have been proposed, including reactions on the surface of dust grains and reactions in the gas-phase. On one hand, \citet{Charnley95} used gas-phase reactions starting with the ethanol (C$_{2}$H$_{5}$OH, hereafter ET) to produce the EF. Recently, the formation of EF has been included in a new gas-phase chemical network \citep{Taquet16}, in which formic acid (HCOOH) and ammonia (NH$_{3}$) favor the formation of EF. On the other hand, \citet{Garrod08} proposed an alternative scenario based on grain-surface chemistry, in which EF can be formed from ET and MF via cosmic ray-induced photodissociation.

The different chemical routes proposed should be checked with observations of EF and its chemically related species. However, since EF is a very complex and hence low abundant molecule, it has been detected so far only towards 4 sources: three massive star forming regions Sgr B2(N) \citep{Belloche09}, W51 e2 \citep{Rivilla17b}, Orion KL \citep{Tercero13, Tercero15}, and one low-mass star forming region Barnard 1b \citep{Marcelino18}. The works by \citet{Belloche09}, \citet{Rivilla17b} and \citet{Tercero13} all used single-dish IRAM 30m data. Single-dish observations give the line intensity integrated over a large beam area and do not provide the spatial distribution of EF nor an accurate value of the source-averaged column density. To better constrain the formation of this species, interferometric observations at high angular resolution and good sensitivity are needed. \citet{Tercero15} provided an ALMA map of a single transition of EF (at 237.567 GHz, with E$_{u}$ = 275.9 K), which peaks mainly towards the compact ridge. However, a more complete multi-transition study involving many EF transitions at different energy levels is needed to: i) determine the excitation conditions of the molecule and thus the column density in the different components of Orion KL, and ii) compare the spatial distribution and abundances of EF with that of the molecules proposed to play a role in its formation.

With this purpose, we study in this paper multiple transitions of EF, covering the spectral range 214-247 GHz, and energy levels from 56 to 788 K, using interferometric ALMA observations with high angular resolution ($\sim$1.$^{\prime\prime}$5). We identified EF, and also the chemically related molecules ET, MF and HCOOH. We studied their relative abundances and relative spatial distributions, and we compared with the predictions of the theoretical gas-phase and grain-surface models, so that we can further understand the chemical routes of EF. This paper is organized as follows. In Section 2, we introduce the observations. The results and derivation of physical parameters are presented in Section 3 and Section 4, respectively. A general discussion is presented in Section 5. We summarize the results in Section 6.

\section{OBSERVATIONS}\label{s:obs}
We used data from the ALMA science verification (SV) program\footnote{The data are available to the public and can be accessed from the ALMA-SV website at https://almascience.nrao.edu/alma-data/science-verification.}. The observations were carried with 16 antennas of 12 m in January 2012. The phase center of the observations is R.A.(J2000)=$05^\mathrm{h}35^\mathrm{m}14^\mathrm{s}.35$ and Decl.(J2000)=$-05^{\circ}22^{'}35^{\prime\prime}.00$. The baseline lengths range from 17 to 265~m. A total of 20 spectral windows of 1.875 GHz completely cover a frequency range from 214 GHz to 247 GHz. Each spectral window consists of 3840~channels with a uniform spectral resolution of 0.448~MHz ($\sim0.7$~km~s$^{-1}$). Callisto and the quasar J0607-085 were used for the flux and phase calibration, respectively. The 1 Jy beam$^{-1}$ correspond to a main beam brightness temperature of $\sim$ 9 K. The primary beam size of each 12~m antenna is about 30$^{\prime\prime}$. The calibrated UV data were exported to Miriad format, and images were using the Miriad software package \citep{Sault95}. The synthesized beam size with natural weighting for spectral images is approximately $1.^{\prime\prime}8\times1.^{\prime\prime}3$ ($\sim$ 745 AU$\times$ 538 AU) with a PA of $-2.6^{\circ}$.

\section{RESULTS}\label{s:res}
\subsection{Line Identification}
We identified spectral line transitions using the eXtended CASA Line Analysis Software Suite (XCLASS; \citealt{Moller17}), which accesses the Cologne Database for Molecular Spectroscopy (CDMS\footnote{http://cdms.de}; M\"{u}ller et al.\ 2001, 2005) and Jet Propulsory Laboratory (JPL\footnote{http://spec.jpl.nasa.gov}; Pickett et al.\ 1998) molecular databases. We used the CDMS entries for EF, ET and t-HCOOH. We used the JPL entry for MF. To identify lines, we set the following constraints \citep{Snyder05,Brouillet15}: the rest frequency must be consistent with the value from laboratory measurement; all detected transitions for a given species must have similar Local standard of rest (LSR) velocity of the source; the transitions at different frequency have different beam dilution effect; for a specific species, transitions having similar excitations condition should have similar spatial distribution. Over the full spectral coverage, we identified 82 (above 5$\sigma$) unblended lines of EF (49 of the \emph{trans} conformer and 33 of the \emph{gauche} conformer) with upper level energies (E$_{\rm u}$) in the range 152--411 K, 469 lines of MF with E$_{\rm u}$ 36--797 K, 204 lines of ET (67 trans and 137 gauche) with E$_{\rm u}$ ranging from 24 to 672 K, and 12 lines of trans-HCOOH with E$_{\rm u}$ of 59--262 K. In this work, all lines below the $5\sigma$ detection level are ignored, where $\sigma$ is the rms noise. In order to compute rms of a certain transition, we measured the rms value considering the line-free channels among the nearby 50 channels. We show in Figure~1 selected transitions of EF that appear unblended or only slightly blended towards the two selected positions (see below): the compact ridge and/or the hot core southwest region. We also present the full observed spectra (black curve) in Figure~A.1, and overplot all the detectable transitions of EF (unblended and blended). The observed spectra of MF and ET are showed in Figure~2 and t-HCOOH is showed in Figure~3. The spectra of EF, MF and ET are extracted from the positions of the intensity peaks of EF (see Figure~4a), which are R.A.=$05^\mathrm{h}35^\mathrm{m}14^\mathrm{s}.430$, Decl.=$-05^{\circ}22^{'}34^{"}.26$ for the hot core-SW and R.A.=$05^\mathrm{h}35^\mathrm{m}14^\mathrm{s}.107$, Decl.=$-05^{\circ}22^{'}36^{"}.84$ for the compact ridge. The spectra of t-HCOOH are extracted from the intensity peaks of t-HCOOH (see Figure~4d, R.A.=$05^\mathrm{h}35^\mathrm{m}14^\mathrm{s}.485$, Decl.=$-05^{\circ}22^{'}32^{"}.56$ for the hot core-SW, and R.A.=$05^\mathrm{h}35^\mathrm{m}14^\mathrm{s}.584$, Decl.=$-05^{\circ}22^{'}27^{"}.46$ for the hot core-N). The rest frequencies, quantum numbers, the line strength (S$\mu^{2}$), and the upper level energies (E$_{\rm u}$) of selected lines of EF are listed in Table A.1. We performed Gaussian fits to every transition. The LSR velocity, full-width half maximun and peak intensities of each transitions are given in columns 5 to 7 (for the hot core-southwest) and 9 to 11 (for the compact ridge). The detection level of each transition is given in column 8 and 12 in Table A.1. The last column of Table A.1 lists the 1$\sigma$ channel rms noise level of each transition. Table~A.2, A.3, A.4 list the unblended lines of MF, ET and formic acid, respectively. From the Figure~1 and A.1, one can see that EF lines are less blended toward the compact ridge than in the hot core-SW, and that the molecular lines have emission intensities slightly weaker toward the compact ridge than toward the hot core-SW.

\subsection{Spatial Distributions}

We show in Figure~4 the line integrated maps of EF, MF, ET and t-HCOOH at different (lower, intermediate and higher) upper level energies, with the aim of evaluating how the molecular emission varies with the energy of the transitions. It should be noted that it is difficult to present the EF integrated maps at low (e.g., E$_{u}$ = 222 K) and high (e.g., E$_{u}$ = 596 K) energies due to contamination from other species, especially toward the hot core-SW. We show three lines at intermediate energy levels (306--320 K) that are completely free from contamination in Figure~4a. In Figure~4, the peak positions of continuum hot core (R.A.=$05^\mathrm{h}35^\mathrm{m}14^\mathrm{s}.543$, Decl.=$-05^{\circ}22^{'}31^{"}.21$) and compact ridge (R.A.=$05^\mathrm{h}35^\mathrm{m}14^\mathrm{s}.064$, Decl.=$-05^{\circ}22^{'}36^{"}.97$) are marked as red crosses. The coordinates of two peak positions are from \citet{Peng17} that used the same dataset.

The emission of EF distributes over the compact ridge and the hot core-SW. The gas map of MF with low-energy level (E$_{u}$ = 236 K) indicates widespread emission and has two main concentrations: the compact ridge and the hot core, in agreement with previous studies (e.g., Favre et al.\ 2011; Widicus Weaver \& Friedel\ 2012; Favre et al.\ 2014; Sakai et al.\ 2015). However, at intermediate (E$_{u}$ = 462 K) and higher (E$_{u}$ = 617 K) energy levels, the emissions of MF becomes more compact and are mainly from the compact ridge and the hot core-SW. We note that the emission of MF at the compact ridge is stronger than at the hot core for low energies, and emission at the hot core becomes stronger at higher upper level energies (e.g., E$_{\rm u}$ = 617 K; see Figure~2 and Figure~4(b)), consistent with the results presented by \citet{Favre11}. This indicates that lines having higher upper level energies are easily excited in hotter environment. A salient result is that the emission of EF and MF mainly come from the hot core-SW and compact ridge, and peak at the same position toward these two regions. Unlike EF and MF, ET has no strong emission in the compact ridge, and it mainly arises from the hot core-SW, which agrees with the previous works (e.g., \citealt{Guelin08,Feng15}).

 The velocity integrated intensity maps of t-HCOOH with three different energies (70--215 K) are shown in Figure~4d. One emission peak of t-HCOOH is mainly at the hot core-SW, while another one is at the north of hot core (hereafter hot core-N). No compact emission of these relatively high energy transitions of HCOOH is detected towards the compact ridge molecular peak traced by EF and MF. We note that previous observations of lower excitation lines (14--70 K) of HCOOH (\citealt{Liu02,Widicus Weaver12,Pagani17}) showed widespread distribution towards the S-SW of the compact ridge. However, this gas component, seen more clearly with poorer spatial resolution (3$^{\prime\prime}$-6$^{\prime\prime}$; \citealt{Liu02,Widicus Weaver12}), do not seem related with the compact gas traced by EF and MF.

\section{DERIVATION OF PHYSICAL PARAMETERS}
\subsection{Rotational temperatures and column Densities}
The combined usage of XCLASS and MAGIX (Modeling and Analysis Generic Interface for eXternal numerical codes; \citealt{Moller13,Moller17}) provides the best fit to the data. We have fitted the spectra extracted from the intensity peaks of EF in the hot core-SW and the compact ridge region, and spectra from the intensity peaks of HCOOH toward the hot core-SW and the hot core-N (see Section 3.1). Under the assumption of local thermodynamical equilibrium (LTE), XCLASS takes into account the beam dilution, the line opacity and line blending \citep{Moller17}, and generates the synthetic spectrum of the different molecular species. The main modeling parameters of XCLASS for each molecule are the source size, the line full width at half maximum $\Delta \upsilon$, the velocity offset with respect to the systemic velocity of the object v$_{\rm off}$, the rotational temperature T$_{\rm rot}$ and the column density N. We have derived the velocity offsets and line widths from Gaussian fits. We have derived the source size by performing two-dimension Gaussian fits to the line images, by using Miriad task IMFIT. We obtain the source sizes of 1.6$^{\prime\prime }$ for the hot core-SW and compact ridge based on the integrated maps of EF. We have fixed the $\Delta \upsilon$ and v$_{\rm off}$. The rotational temperature and the column density are left as free parameters. These two parameters produced by XCLASS are used as initial guesses, and then MAGIX is used to optimize the fit and find the best solutions of T$_{\rm rot}$ and N, and provide corresponding error estimates. The algorithms we adopted are the Genetic, Levenberg-Marquardt and Error estimation using Interval-Nested-Sampling.

We stress that in the case of the conformers of EF and ET, we are able to fit both conformers using the same physical parameters. So both conformers of EF and ET are fitted together. We show the LTE model spectra overlaid on the observed spectra in Figure~1 (EF) and Figure~2 (MF and ET) toward the hot core-SW and the compact ridge. Figure~3 shows the LTE spectra of t-HCOOH toward the hot core-SW and the hot core-N. The derived parameters from two positions, including rotational temperature $T_{\rm rot}$, total column density N and abundance $\chi$ are listed in Table~1. See more details about abundances in section 4.2. In the following we comment individually the rotational temperature and column density obtained from each species:

\textbf{EF (C$_{2}$H$_{5}$OCHO)}: The detected unblended transitions cover a large E$_{\rm u}$ range (152--411 K), which allow to constrain well the rotational temperature, and then the column density. We have obtained T$_{\rm rot}$ = 122 $\pm$ 34 K, and a column density of N = (9.0 $\pm$ 3.0) $\times$ 10$^{15}$ cm$^{-2}$ for the hot core-SW, and T$_{\rm rot}$ = 103 $\pm$ 13 K, N = (6.0 $\pm$ 3.0) $\times$ 10$^{15}$ cm$^{-2}$ for the compact ridge. Meanwhile, for the purpose of comparison with other works, we list the number of detected lines, corresponding E$_{\rm u}$ range, observations information and source size they used in Table~2. The temperature T$_{\rm rot}$ we obtained is similar to that derived in W51 (78 $\pm$ 10 K) by \cite{Rivilla17b}, and to that assumed in Orion KL compact ridge (100 $\pm$ 20 K) by \citet{Tercero15} and Sgr B2(N) (100 K) by \citet{Belloche09}. The column density we obtained in the compact ridge is higher to that of \citet{Tercero15}. They assumed T$_{\rm rot}$ of 100 K and $\theta_{\rm s}$ of 3$^{\prime\prime}$ for one single transition to obtain column density N of (0.2 $\pm$ 0.04) $\times$ 10$^{16}$ cm$^{-2}$. This value would be 0.7$\times$ 10$^{16}$ cm$^{-2}$, similar to our value, if we consider the same source size of 1.6$^{\prime\prime}$.

As mentioned before, multiple unblended or slightly blended lines of the both conformers of EF are clearly identified, as shown in Figure~1. We note that a few transitions (e.g., 218.74259 GHz, 218.74470 GHz, 220.00368 GHz, 223.89863 GHz, 224.28239 GHz) are weaker than the LTE fit. A possible explanation is a poor baseline subtraction at these spectral ranges, since Orion is a line-rich source, and it is difficult to find free-line channels, and then the continuum is not perfectly substracted. On the other hand, some of the observed lines might be weaker than the LTE fit due to the presence of nearby absorption lines from other species (e.g. 218.752 GHz, 219.658 GHz and 225.240 GHz).
There are some lines, at 219.40243 GHz, 219.65815 GHz, 219.78133 GHz, 219.78771 GHz and 220.17206 GHz, for which the LTE fit predicts higher line intensities than observed. We found that the spectroscopic parameters (in particular S$\mu^{2}$) of these five transitions show discrepancies between the CDMS and the JPL entries. Other lines that have similar spectroscopic parameters in two database have good fitting results. We suspect that there are some issues affecting the spectroscopic information of some of the transitions with discrepancies.

\textbf{MF (CH$_{3}$OCHO)}: We note that the stronger lines are overestimated toward the hot core-SW, this could be due to line opacity effects. These stronger lines have high S$\mu^{2}$. If we leave the source size as a free parameter, all lines of MF can be fitted better, including high S$\mu^{2}$ lines, which appears saturated with optical depth of $\sim$4. The strong lines are expected to have more extended gas distribution, and interferometric observations of ALMA will partially filter out diffuse gas component. This is likely to cause decreasing intensities of strong lines. The best fit parameters of MF are T$_{\rm rot}$ = 135 $\pm$ 5 K and N = (7 $\pm$ 3) $\times$ 10$^{17}$ cm$^{-2}$ toward the hot core-SW, and 93 $\pm$ 5 K and (9 $\pm$ 4) $\times$ 10$^{17}$ cm$^{-2}$ for the compact ridge. Our derived rotational temperatures in the hot core-SW and compact ridge are mostly consistent with the other works, but column densities in the two regions are larger than the values provided by others (see Table~2, e.g., \citealt{Favre11,Feng15,Tercero15}). Our column density is higher than results from \citet{Favre11} by a factor of $\sim$4--5 probably because they cover lower E$_{\rm u}$ transitions, especially in the compact ridge. The lower column densities found by \citet{Feng15} is easily explained because of their large beam (5.$^{\prime\prime}$0$\times$4.$^{\prime\prime}$0), which also includes less dense gas. The difference of column density between our work and \citet{Tercero15} can be easily explained due to the different source size used. If one consider $\theta_{\rm s}$ of 1.6$^{\prime\prime}$ instead of 3$^{\prime\prime}$, the \citet{Tercero15} estimate would be 84 $\times$ 10$^{16}$ cm$^{-2}$, in remarkably agreement with our results.

\textbf{ET (C$_{2}$H$_{5}$OH)}: Regarding ET, we have found significant discrepancies in the spectroscopic parameters between the CDMS and JPL entries affecting some transitions, e.g., those at 213.760 and 213.856 GHz. Comparing the LTE spectra of both databases, we have seen that the CDMS entry is more consistent with the observed spectrum, so that we have used it to derive the physical properties. The LTE fitting gives T$_{\rm rot}$ = 144 $\pm$ 13 K and N = (1.3 $\pm$ 0.5)$\times$ 10$^{17}$ cm$^{-2}$ toward the hot core-SW, and 111 $\pm$ 15 K and (2.0 $\pm$ 0.5) $\times$ 10$^{16}$ cm$^{-2}$ for the compact ridge. The column densities derived from our work are higher than those reported in \citet{Feng15} (see Table 2), owing to our higher angular resolution. Again for result from \citet{Tercero15}, if one consider $\theta_{\rm s}$ =1.6$^{\prime\prime}$, the column density would be 21 $\times$ 10$^{16}$ cm$^{-2}$, which is of the same order than our value (less than a factor of 2).

\textbf{Formic acid (t-HCOOH)}: We derive T$_{\rm rot}$ = 120 $\pm$ 16 K and N = (6.2 $\pm$ 1.8) $\times$ 10$^{15}$ cm$^{-2}$ toward the hot core-SW peak, and 106 $\pm$ 27 K and (2.8 $\pm$ 0.8) $\times$ 10$^{15}$ cm$^{-2}$ toward the hot core-N.

We give in Table 3 the ratios of [MF/EF] and [MF/ET] from our work and a few other works toward the hot core-SW and the compact ridge. The [MF/EF] of 140$\pm$81 toward the compact ridge in our work is reasonable agreement with the ratio of 120 in \citet{Tercero15}. Toward the hot core-SW, the [MF/ET] of 6$\pm$3 is in agreement with that of 4$\pm$3 in \citet{Feng15}, and toward the compact ridge our result of 42$\pm$24 is similar to those from \citet{Feng15} (17$\pm$12) and \citet{Crockett14} (20) if considering the uncertainties.

\subsection{Molecular abundances}
The fractional abundance of a certain molecule relative to H$_{2}$ depends on the column density of the certain molecule, N, and that of the H$_{2}$ molecule ($\chi$ = N/N$_{\rm H_{2}}$). The estimations of N$_{\rm H_{2}}$ for each Orion KL component are presented in many works (e.g., Tercero et al. 2010; Crockett et al. 2014; Favre et al. 2011; Feng et al. 2015), and N$_{\rm H_{2}}$ varies in every work because of different spatial resolutions. We used the dust continuum emission of the ALMA observations to estimate the column density of H$_{2}$. In the assumption of optically thin emission from dust, and taking a grain radius of 0.1 $\mu$m, a grain density of 3 g cm$^{-3}$, and a gas-to-dust ratio of 100, the column density of H$_{2}$ can be derived by the formula (e.g., \citealt{Lis91,Qin10}):
\begin{equation}\label{1}
  N_{H_{2}} = 8.1 \times 10^{17} \frac{e^{h\nu/\kappa T_{d}}-1}{Q_{\nu}\Omega}\left(\frac{S_{\nu}}{Jy}\right)\left(\frac{\nu}{GHz}\right)^{-3} (cm^{-2}),
\end{equation}

where T$_{d}$ is the dust temperature, Q$_{\nu}$ is the grain emissivity at frequency $\nu$, S$_{\nu}$ is the peak intensity of the continuum and $\Omega$ is the beam solid angle. We assumed that the dust temperature equals to the rotation temperature of the core. We used dust temperature of 200 K for the hot core and 100 K for the compact ridge (e.g., Wu et al. 2014; Favre et al. 2011), and grain emissivity of 2 $\times$ 10$^{-5}$ at 230 GHz \citep{Lis91}. We adopt S$_{\nu}$ of 1.34 $\pm$ 0.06 Jy~beam$^{-1}$ for the hot core and 0.28 $\pm$ 0.03 Jy~beam$^{-1}$ for the compact ridge from \citet{Wu14} that used same ALMA data. Therefore we obtain H$_{2}$ column densities of (4.7 $\pm$ 0.2) $\times$ 10$^{24}$ cm$^{-2}$ and (2.0 $\pm$ 0.2) $\times$ 10$^{24}$ cm$^{-2}$ for the hot core and the compact ridge, respectively. We list the abundances relative to H$_{2}$ of EF, MF, ET and t-HCOOH toward the two positions in Table~1. In comparison with abundances for each species from other works (see Table~2), one can see that:

1) The abundance of EF towards the hot core and the compact ridge are very similar, 0.19 $\times$ 10$^{-8}$ and 0.3 $\times$ 10$^{-8}$ respectively. This value is in good agreement with the ones derived in the other two massive star forming regions where EF has been detected: 1 $\times$ 10$^{-8}$ in W51 \citep{Rivilla17b} and 0.36 $\times$ 10$^{-8}$ on Sgr B2(N) \citep{Belloche09}.

2) The abundances of ET and MF we have derived are in better agreement with those from \citet{Feng15}, compared with the column densities. This is because the effect of the different beams are compensated by dividing by N$_{\rm H_{2}}$ (namely, their N is lower but also their N$_{\rm H_{2}}$).

3) The abundances of MF from \citet{Favre11} are lower, especially in the compact ridge (3.1 $\times$ 10$^{-8}$). This could be because the range of energies of the lines used in the compact ridge by \citet{Favre11} is much lower than the one we have used (see Table~2), and thus they are tracing less dense gas.

4) Our derived abundance of MF ((42$\pm$20) $\times$ 10$^{-8}$) in the compact ridge agrees very well with those derived by \citet{Gong15} ((47$\pm$6) $\times$ 10$^{-8}$) and \citet{Crockett14} (33 $\times$ 10$^{-8}$).

We compare graphically the abundances (Figure~5a) and rotational temperatures (Figure~5b) of each molecule derived in different regions (Orion KL, W51 and Sgr B2(N)) with the theoretical values from the grain-surface model by \citet{Garrod08} and gas-phase model by \citet{Taquet16}. In the next section we will discuss which of the two proposed chemistries is in better agreement with the observations in star-forming regions.

\section{Discussion}
 We have shown that both conformers of EF can be well fitted simultaneously, and hence that they share the same abundance. The relative abundance of the two conformers is $N_{\rm g}$/$N_{\rm t}$ = 2e$^{-\Delta E/T_{\rm kin}}$ \citep{Rivilla17b}, where $T_{\rm kin}$ is kinetic temperature of the gas and $\Delta E$ is energy difference between the two ground states of two conformers. $\Delta E$ is in the range of 64--124 K \citep{Riveros67}, which implies a $T_{\rm kin}$ in the range 93--178 K when considering $N_{\rm g}$/$N_{\rm t}$ = 1. The $T_{\rm kin}$ would be equal to the rotational temperature T$_{\rm rot}$, if we assume that EF is thermalized. The T$_{\rm rot}$ derived in our work (122 $\pm$ 34 K for hot core-SW and 103 $\pm$ 13 K for compact ridge, see Table~1) are compatible to the range of 93--178 K, which suggests that EF is indeed thermalized.

As we stated above, three molecules (MF, ET and HCOOH) may play a important role in forming EF, so it is necessary to take these related species into account to study the proposed chemical routes to EF. Table~1 and Figure~5b showed that $T_{\rm rot}$ in the hot core are in the range of 120--144 K, and 93--111 K in the compact ridge, indicating that these four molecules share similar excitation conditions, and hence they are expected to trace similar gas components. This means that if there is no emission in some region (for instance ET in the compact ridge), this is due to a chemical effect.

Two main pathways have been proposed to form EF, one based on gas-phase chemistry and other based on chemistry on the surface of dust grains. We compare in the following the result found in star-forming regions (Orion KL, W51 and Sgr B2(N)) with the predictions of the proposed models.

 1) Gas-phase formation. Based on the model from \citet{Charnley95}, their gas-phase route is able to produce EF abundance of $\sim$10$^{-10}$ in 10$^{4}$-10$^{5}$ yr (C$_{2}$H$_{5}$OH + H$^{+}$ $\rightarrow$ C$_{2}$H$_{5}$OH$^{+}_{2}$, C$_{2}$H$_{5}$OH$^{+}_{2}$ + H$_{2}$CO $\rightarrow$ HC(OH)OC$_{2}$H$^{+}_5$ + H$_{2}$, HC(OH)OC$_{2}$H$^{+}_5$ + e$^{-}$ $\rightarrow$ C$_{2}$H$_{5}$OCHO + H). The observed EF abundances of 0.2 $\times$ 10$^{-8}$ (hot core) and 0.3 $\times$ 10$^{-8}$ (compact ridge) (see Table~1) are much larger than that predicted by gas-phase route. The estimated abundances in W51 (1.0 $\times$ 10$^{-8}$) and Sgr B2(N) (3.6 $\times$ 10$^{-9}$) are also larger than that predicted ones. So this gas-phase reaction can not produce enough abundance of EF. A new gas-phase network for the formation of COMs have been proposed by \citet{Taquet16}. Considering proton transfer (PT) reaction with HCOOH and NH$_{3}$ (C$_{2}$H$_{5}$OH$^{+}_{2}$ + HCOOH $\rightarrow$ HC(OH)OC$_{2}$H$^{+}_5$ + H$_{2}$O, HC(OH)OC$_{2}$H$^{+}_5$ + NH$_{3}$ $\rightarrow$ C$_{2}$H$_{5}$OCHO +NH$^{+}_{4}$), their model is able to produce the observed EF abundances of 10$^{-8}$--10$^{-7}$ in a reasonable timescale for hot molecular cores of 10$^{4}$ $-$ 10$^{5}$ yr. In the hot core-SW, our derived HCOOH abundance of (1.3$\pm$0.4) $\times$ 10$^{-9}$ is much lower than those required for gas-phase model (see panel a(4) of Figure~5). Moreover, the ALMA maps show that there is no similarity in spatial distributions and emission peaks between HCOOH and EF (see Figure~4).

Regarding the distribution of NH$_{3}$ (transitions with E$_{\rm u}$ $>$ 408 K) detected by \citet{Goddi11}, it peaks towards the hot core and IRC7 regions, while no emission is detected towards the compact ridge peak traced by EF and MF (see their Fig. 2). Since both the hydrogen column densities and rotational temperatures found towards the two positions studied in our work, hot core-SW (N$_{\rm H_{2}}$ = (4.7 $\pm$ 0.2) $\times$ 10$^{24}$ cm$^{-2}$, T$_{\rm rot}$ = 120--144 K) and compact ridge (N$_{\rm H_{2}}$ = (2.0 $\pm$ 0.2) $\times$ 10$^{24}$ cm$^{-2}$, T$_{\rm rot}$ = 93--111 K), are relatively similar, the clear presence of NH$_{3}$ in the hot core region and the lack in the compact ridge peak, point to a chemical differentiation and not to excitation effects. Therefore, similarly to HCOOH, there is no spatial coincidence between NH$_{3}$ and EF. In the hot core region, \citet{Goddi11} derived a column density of NH$_{3}$ of (1-4) $\times$ 10$^{17}$ cm$^{-2}$, which implies an abundance of (2$-$9) $\times$ 10$^{-8}$. This value is lower than the gas model value by more than an order of magnitude. In summary, the molecular abundances and spatial distributions of HCOOH and NH$_{3}$ indicate that the proposed gas-phase routes are not able to produce the observed EF abundance in Orion KL, as also noted by \citet{Rivilla17b} in W51.

2) Grain-surface formation. According to the coupled gas-phase and grain-surface chemical network by \citet{Garrod08}, EF can be formed on grain surfaces by the reactions between radicals CH$_{3}$ and CH$_{2}$OCHO or between HCO and C$_{2}$H$_{5}$O. MF is the primary source of precursor radical of EF (CH$_{3}$OCHO + h$\nu$ $\rightarrow$ CH$_{2}$OCHO, CH$_{2}$OCHO + CH$_{3}$ $\rightarrow$ C$_{2}$H$_{5}$OCHO). When MF evaporates, ET (C$_{2}$H$_{5}$OH) becomes dominant in forming EF ( C$_{2}$H$_{5}$OH + h$\nu$ $\rightarrow$ C$_{2}$H$_{5}$O; C$_{2}$H$_{5}$O + HCO $\rightarrow$ C$_{2}$H$_{5}$OCHO). The predicted abundance of EF is 2.3 $\times$ 10$^{-9}$ in this model (see \citealt{Belloche09}), which is in reasonably agreement with the range of observed values in our work ((2$-$3)$\times$10$^{-9}$) and in W51 e2 and Sgr B2(N) ((3.6$-$10)$\times$10$^{-9}$). In addition, the predicted molecular ratio MF/EF is 72/1 (in Basic model, see Table~15 of \citealt{Belloche09}), which is close to our observed ratio values of (79$\pm$36)/1 for the hot core and (140$\pm$81)/1 for the compact ridge. The excitation temperature of EF (122$\pm$34 K for the hot core and 103$\pm$13 K for the compact ridge) and MF (135$\pm$5 K for the hot core and 93$\pm$5 K for the compact ridge) derived in Orion KL are also close to the temperature associated with the peak gas-phase temperatures of \citet{Garrod08} model (110 K for EF and 81 K for MF) (see Figure~5b). The abundances of ET derived from our fitting ((2.7$\pm$1.0) $\times$ 10$^{-8}$ for the hot core and (1.0$\pm$0.3) $\times$ 10$^{-8}$ for the compact ridge), 7.9$\times$10$^{-8}$ in W51 and 5.7$\times$10$^{-8}$ in Sgr B2(N) (see Table~2 and Figure~5a) are lower than abundance from \citet{Garrod08} model (1.0$\times$10$^{-7}$). This may indicate that the dominant route to the formation of EF is grain surface chemistry from MF and not from ET. This is also supported by the spatial distribution of the different molecules revealed by the ALMA observations (see Figure~4): EF and MF share the same intensity peaks toward the hot core-SW and the compact ridge, while there is no strong ET emission towards the compact ridge. Considering all this together, EF is likely formed on grains by addition of CH$_{3}$ to functional-group radicals (CH$_{2}$OCHO) derived from CH$_{3}$OCHO and then evaporated to the gas phase.
\section{Summary}
As the the next step in chemical complexity of esters, ethyl formate (C$_{2}$H$_{5}$OCHO) is a low abundant molecule and it is detected so far only toward three sources. The formation routes of EF are still poorly known. Multiple transitions of EF covering a wide range of energy levels are needed to explore it's excitation conditions and formation routes. The different chemical routes proposed should be checked with observations of EF and its chemically related species (methyl formate (CH$_{3}$OCHO), ethanol (C$_{2}$H$_{5}$OH) and trans-HCOOH). We have identified and imaged multiple lines of EF, MF, ET and t-HCOOH observed with ALMA at band 6 (214--247 GHz). The identified transitions of three molecules span a wide range of upper-level energy so that we can perform a detailed LTE analysis to derive accurate physical parameters. We have compared our results with the two main chemical routes proposed to the formation of EF (gas-phase chemistry and grain surface chemistry). The main findings are as follows:

1. We have identified 82 unblended lines of EF (49 of the trans conformer and 33 of the gauche conformer), 469 lines of MF, 204 lines of ET (67 of the trans conformer and 137 of the gauche conformer) and 12 lines of t-HCOOH over the full ALMA band 6.

2. The spatial distribution of EF exhibits two main emission regions (compact ridge and hot core-SW). The gas emission of MF is very similar to that of EF, while ET mainly originates from the hot core-SW but not from the compact ridge. t-HCOOH mainly originate from the south-west of hot core and hot core-N.

3. The derived rotational temperatures and column densities of EF are 122 $\pm$ 34 K, (9.0 $\pm$ 3.0) $\times$ 10$^{15}$ cm$^{-2}$ for the hot core-SW, and 103 $\pm$ 13 K, (6.0 $\pm$ 3.0) $\times$ 10$^{15}$ cm$^{-2}$ for the compact ridge. We also derived the best fit parameters of chemically related molecules MF, ET and t-HCOOH.

4. We have compared our results with the two proposed routes to form EF, based on gas-phase chemistry \citep{Taquet16} and grain surface chemistry \citep{Garrod08,Belloche09}. The observed abundances of NH$_{3}$ and HCOOH are much lower than those required for gas-phase model to produce the observed abundances of EF. Moreover, there is no similarity in spatial distributions and emission peaks among HCOOH, NH$_{3}$ and EF. So these gas-phase routes may not produce the observed EF abundance in Orion KL and other star-forming regions. The derived EF abundance of (2-3)$\times$10$^{-9}$ is in good agreement with the one predicted by the grain surface model. The observed ratios of MF/EF ((79$\pm$36)/1 for the hot core and (140$\pm$81)/1 for the compact ridge are also in good agreement with the grain-surface model. In addition, the gas emissions of EF and MF peak at same position in the hot core and compact ridge, while ET spatial distribution is different, lacking in the compact ridge. Altogether, these results suggest that EF is likely formed on grains by addition of CH$_{3}$ to functional-group radicals (CH$_{2}$OCHO) derived from CH$_{3}$OCHO and then evaporated to the gas phase.

\bigskip

\acknowledgements This paper
makes use of the following ALMA data: ADS/JAO.ALMA\#
2011.0.00009.SV. ALMA is a partnership of ESO (representing
its member states), NSF (USA), and NINS (Japan), together with
NRC (Canada) and NSC and ASIAA (Taiwan), in cooperation
with the Republic of Chile. The Joint ALMA Observatory is
operated by ESO, AUI/NRAO, and NAOJ. We thank Sheng-Li Qin for carefully reading the paper and providing useful comments and suggestions. This work has been supported by the National Natural Science
Foundation of China under grant No. 11433004. V.M. R. has received funding from the European Union's Horizon 2020 research and innovation programme under the Marie Sklodowska-Curie grant agreement No 664931, and from the Italian  Ministero  dell'Istruzione, Universit\`a e Ricerca through the grant Progetti Premiali 2012 - iALMA (CUP C52I13000140001). JX acknowledges support from FONDECYT grant 3170768 and the "Light of West China" program of Chinese Academy of Sciences (CAS).

\clearpage

\begin{deluxetable} {cccccccccccccccccccccc}
\tabletypesize{\tiny}
\tablewidth{-3pc}
\tablenum{1}
\tablecolumns{10}
\tablecaption{Model fitting results of EF, MF, ET and t-HCOOH. }
\tablehead
{ 
\colhead{ Molecular name} &\multicolumn{4}{c}{hot core-SW} & & \multicolumn{4}{c}{compact ridge}\\
\cline{2-6} \cline{8-12} \\
& \colhead{$v_{\rm LSR}$ \tablenotemark{a}} & \colhead{ $\Delta \upsilon$ \tablenotemark{a}} &\colhead{T$_{\rm rot}$}& \colhead{N}& \colhead{$\chi$}& & \colhead{$v_{\rm LSR}$ \tablenotemark{a}} & \colhead{ $\Delta \upsilon$ \tablenotemark{a}} &\colhead{T$_{\rm rot}$}& \colhead{N}& \colhead{$\chi$}& \\
&\colhead{$(km \ s^{-1})$} & \colhead{$(km \ s^{-1})$}&\colhead{($\rm K$)}&\colhead{($10^{16}cm^{-2}$)}& \colhead{($10^{-8}$)}&&\colhead{$(km \ s^{-1})$} & \colhead{$(km \ s^{-1})$}&\colhead{($\rm K$)}&\colhead{($10^{16}cm^{-2}$)} & \colhead{($10^{-8}$)} }

\startdata

 C$_{2}$H$_{5}$OCHO & 6.8$\pm$0.1 & 3.0$\pm$0.1 & 122$\pm$34 & 0.9$\pm$0.3 & 0.19$\pm$0.06& &
                           7.2$\pm$0.2 & 2.2$\pm$0.1 & 103$\pm$13 & 0.6$\pm$0.3   & 0.3$\pm$0.1
                            \\
  &&&&&&&&\\

CH$_{3}$OCHO   & 7.3$\pm$0.1 & 3.2$\pm$0.1 & 135$\pm$5 & 70.0$\pm$26.0  & 15$\pm$5 &
               & 7.1$\pm$0.1 & 2.5$\pm$0.2 & 93$\pm$5 & 85.0$\pm$40.0  & 42$\pm$20
               \\
 &&&&&&&\\
 C$_{2}$H$_{5}$OH  & 7.1$\pm$0.1 & 3.5$\pm$0.1 & 144$\pm$13 & 12.9$\pm$4.8 & 2.7$\pm$1.0   &
                            & 7.7$\pm$0.1 & 2.5$\pm$0.1 & 111$\pm$15 &2.0$\pm$0.5 & 1.0$\pm$0.3
                             \\
  &&&&&&&\\
t-HCOOH  & 6.7$\pm$0.1 & 4.0$\pm$0.2 & 120$\pm$16 & 0.6$\pm0.2$ & 0.13$\pm$0.04 &
                            & $\cdots$ & $\cdots$ &$\cdots$ &$\cdots$
                            \\

\enddata
\tablenotetext{a}{The $v_{\rm LSR}$ and $\Delta \upsilon$ toward two positions are the average values from the Gaussian fitting of several completely unblended lines.}
\end{deluxetable}

\begin{deluxetable} {cccccccccccccccccccccc}
\tabletypesize{\tiny}
\tablewidth{-3pc}
\tablenum{2}
\tablecolumns{6}
\tablecaption{Parameters of EF, MF and ET from best-fit LTE analysis in three sources (Orion KL, W51 and Sgr B2(N)).}
\tablehead
{ 
\colhead{ Species} &\colhead{Source} & \colhead{Num.$^{1}$}  & \colhead{E$_{\rm u}$ range} & \colhead{Obs.$^{2}$}& \colhead{$\theta_{\rm s}$$^{3}$}& \colhead{T$_{\rm rot}$}  & \colhead{N} & \colhead{$\chi$}  & \colhead{reference}\\
& & & ($\rm K$) & &\colhead{($^{\prime\prime}$)} &\colhead{($\rm K$)}&\colhead{($10^{16}cm^{-2}$)}& \colhead{($10^{-8}$)} }

\startdata

 & Orion KL (hc) & 34 & 152$-$387  & (ALMA)1.$^{\prime\prime}$8$\times$1.$^{\prime\prime}$3 & 1.6 & 122$\pm$34 & 0.9$\pm$0.3 &  0.19$\pm$0.06  & \colhead{this work}\\
                    & Orion KL (cr) & 51 & 170$-$411  & (ALMA)1.$^{\prime\prime}$8$\times$1.$^{\prime\prime}$3 & 1.6 & 103$\pm$13 & 0.6$\pm$0.3 &  0.3$\pm$0.1  & \colhead{this work}\\
 EF                    & Orion KL (cr) & 1 & 276$-$276 & (ALMA)2.$^{\prime\prime}$0$\times$1.$^{\prime\prime}$5 & 3.0 & 100$\pm$20 & 0.2$\pm$0.04 & $\cdots$ & \citet{Tercero15}\\
                    & Orion KL (cr) & 90 & $\cdots$ & (IRAM 30m)11$^{\prime\prime}$$-$29$^{\prime\prime}$ & 15.0 & 150$\pm$20 & 0.045$\pm$0.01 & $\cdots$ & \citet{Tercero13}\\
                    & W51 & 46 & 42$-$213 & (IRAM 30m)10$^{\prime\prime}$$-$27$^{\prime\prime}$ & 2.4 & 78$\pm$10 & 2.0$\pm$0.3 & 1.0& \citet{Rivilla17b} \\
                    & Sgr B2(N) & 46 & 34$-$253 & (IRAM 30m)22$^{\prime\prime}$$-$29$^{\prime\prime}$ & 3.0 & 100 & 5.4 & 0.36 & \citet{Belloche09}\\

 \hline
      & Orion KL (hc) & 350 & 36$-$797 & (ALMA)1.$^{\prime\prime}$8$\times$1.$^{\prime\prime}$3 & 1.6 & 135$\pm$5 & 70.0$\pm$26.0 &  15$\pm$5 & \colhead{this work}\\
                  & Orion KL (hc) & 20 & 57$-$312 & (SMA+IRAM 30m)5.$^{\prime\prime}$0$\times$4.$^{\prime\prime}$0 &$\cdots$& 126$\pm$13 & 2.7$\pm$1.1 & 2$\pm$1 & \citet{Feng15}\\
                  & Orion KL (hc) & 14 & 36$-$618 & (PdBI)1.$^{\prime\prime}$79$\times$0.$^{\prime\prime}$79 &$\cdots$& 128$\pm$9 & 16.0$\pm$2.0 & 5.2 & \citet{Favre11}\\
                  & Orion KL (cr) & 458 & 23$-$797 & (ALMA)1.$^{\prime\prime}$8$\times$1.$^{\prime\prime}$3 & 1.6 &  93$\pm$5 & 85.1$\pm$40.0 & 42$\pm$20 & \colhead{this work}\\
 MF                 & Orion KL (cr) & 1 & 122$-$122 & (ALMA)2.$^{\prime\prime}$0$\times$1.$^{\prime\prime}$5 & 3.0 & 100$\pm$20 & 24.0$\pm$5.0 & $\cdots$ & \citet{Tercero15}\\
                  & Orion KL (cr) & 27 & 57$-$312 & (SMA+IRAM 30m)5.$^{\prime\prime}$0$\times$4.$^{\prime\prime}$0 & $\cdots$&  102$\pm$5 & 9.7$\pm$2.1 & 15$\pm$3 & \citet{Feng15}\\
                  & Orion KL (cr) & 12 & 2$-$143 & (Effelsberg-100m)35$^{\prime\prime}$$-$50$^{\prime\prime}$ & 10 & 92$\pm$14 & 18$\pm$2 & 47$\pm$6 & \citet{Gong15}\\
                  & Orion KL (cr) & $\cdots$ & $\cdots$ &(HIFI, IRAM 30m) & 10 &  110 & 13 & 33 & \citet{Crockett14}\\
                  & Orion KL (cr)\tablenotemark{a} & 20 & 36$-$366 & (PdBI)1.$^{\prime\prime}$79$\times$0.$^{\prime\prime}$79 &$\cdots$& 79$\pm$2 & 16.0$\pm$1.0 & 3.1 & \citet{Favre11}\\
                  & W51 & 38 & 28$-$365 & (IRAM 30m)10$^{\prime\prime}$$-$27$^{\prime\prime}$ & 2.4  & 112$\pm$3 & 50.1$\pm$0.6 & 25& \citet{Rivilla17b}\\
                  & Sgr B2(N) & $\cdots$ & $\cdots$ & (IRAM 30m)22$^{\prime\prime}$$-$29$^{\prime\prime}$ &4.0& 80 & 45 & 3.5 & \citet{Belloche09}\\

                   \hline
   & Orion KL (hc) & 186 & 24$-$672 & (ALMA)1.$^{\prime\prime}$8$\times$1.$^{\prime\prime}$3 & 1.6 & 144$\pm$13 & 12.9$\pm$4.8 &  2.7$\pm$1.0 & \colhead{this work}\\
                  & Orion KL (hc) & 5 & 86$-$263 & (SMA+IRAM 30m)5.$^{\prime\prime}$0$\times$4.$^{\prime\prime}$0 &$\cdots$& 137$\pm$33 & 0.8$\pm$0.5 & 0.6$\pm$0.5 & \citet{Feng15}\\
                  & Orion KL (hc) & 1 & 316$-$316 & (ALMA)2.$^{\prime\prime}$0$\times$1.$^{\prime\prime}$5 & 3.0 & 100$\pm$20 & 6.0 & $\cdots$ & \citet{Tercero15}\\
 ET                 & Orion KL (cr) & 162  & 24$-$548 & (ALMA)1.$^{\prime\prime}$8$\times$1.$^{\prime\prime}$3 & 1.6 & 127$\pm$17 & 2.0$\pm$0.5 &  1.0$\pm$0.3 & \colhead{this work}\\
                  & Orion KL (cr) & 5 & 86$-$263 & (SMA+IRAM 30m)5.$^{\prime\prime}$0$\times$4.$^{\prime\prime}$0 & $\cdots$ & 68$\pm$11 & 0.6 & 0.9$\pm$0.5 & \citet{Feng15}\\
                  & Orion KL (cr) & $\cdots$ & $\cdots$ &(HIFI, IRAM 30m) & 10 & 110 & 0.7 & 1.7 & \citet{Crockett14}\\
                  & W51 & 14 & 35$-$327 & (IRAM 30m)10$^{\prime\prime}$$-$27$^{\prime\prime}$ & 2.4 & 95$\pm$3 & 15.8$\pm$0.4 & 7.9 & \citet{Rivilla17b}\\
                  & Sgr B2(N) & $\cdots$ & $\cdots$ & (IRAM 30m)22$^{\prime\prime}$$-$29$^{\prime\prime}$ & 3.0 & 100 & 84 & 5.7 & \citet{Belloche09}\\

\enddata
\tablenotetext{a}{toward the 7.6 km~s$^{-1}$ component for the compact ridge}
\tablecomments{ "hc" and "cr" indicate  hot core and compact ridge, respectively.
              $^{1}$ Number of identified lines. $^{2}$ information of observations, including telescope name, synthesized beam size for interferometric observations and HPBW beam size for single-dish observations. $^{3}$ source size for modeling. The "$\cdots$" in $\theta_{\rm s}$ means that they derived column density and rotational temperature by rotation temperature diagram (RTD) method.}
\end{deluxetable}

\begin{deluxetable} {cccccccccccccccccccccc}
\tabletypesize{\scriptsize}
\tablewidth{-3pc}
\tablenum{3}
\tablecolumns{7}
\tablecaption{Molecular column density  ratios toward Orion KL in different works.}
\tablehead
{   &\multicolumn{2}{c}{[MF/EF]}&  &\multicolumn{3}{c}{[MF/ET]} \\
\cline{2-3}  \cline{5-7}\\
    &\colhead{this work} &\colhead{\citet{Tercero15}}&  &\colhead{this work} &\colhead{\citet{Feng15}}&\colhead{\citet{Crockett14}}
}

\startdata

\colhead{hot core-SW} & 79$\pm$36 & $-$ &  & 6$\pm$3 & 4$\pm$3 & $-$ \\
\hline
\colhead{compact ridge} &  140$\pm$81 & $\simeq 120$ &  & 42$\pm$24 & 17$\pm$12 & 20\\

\enddata
\end{deluxetable}





\clearpage
\begin{figure*}
\newcounter{1}
\setcounter{1}{\value{figure}}
\setcounter{figure}{0}
\centerline{\resizebox{1.0\hsize}{!}{\includegraphics{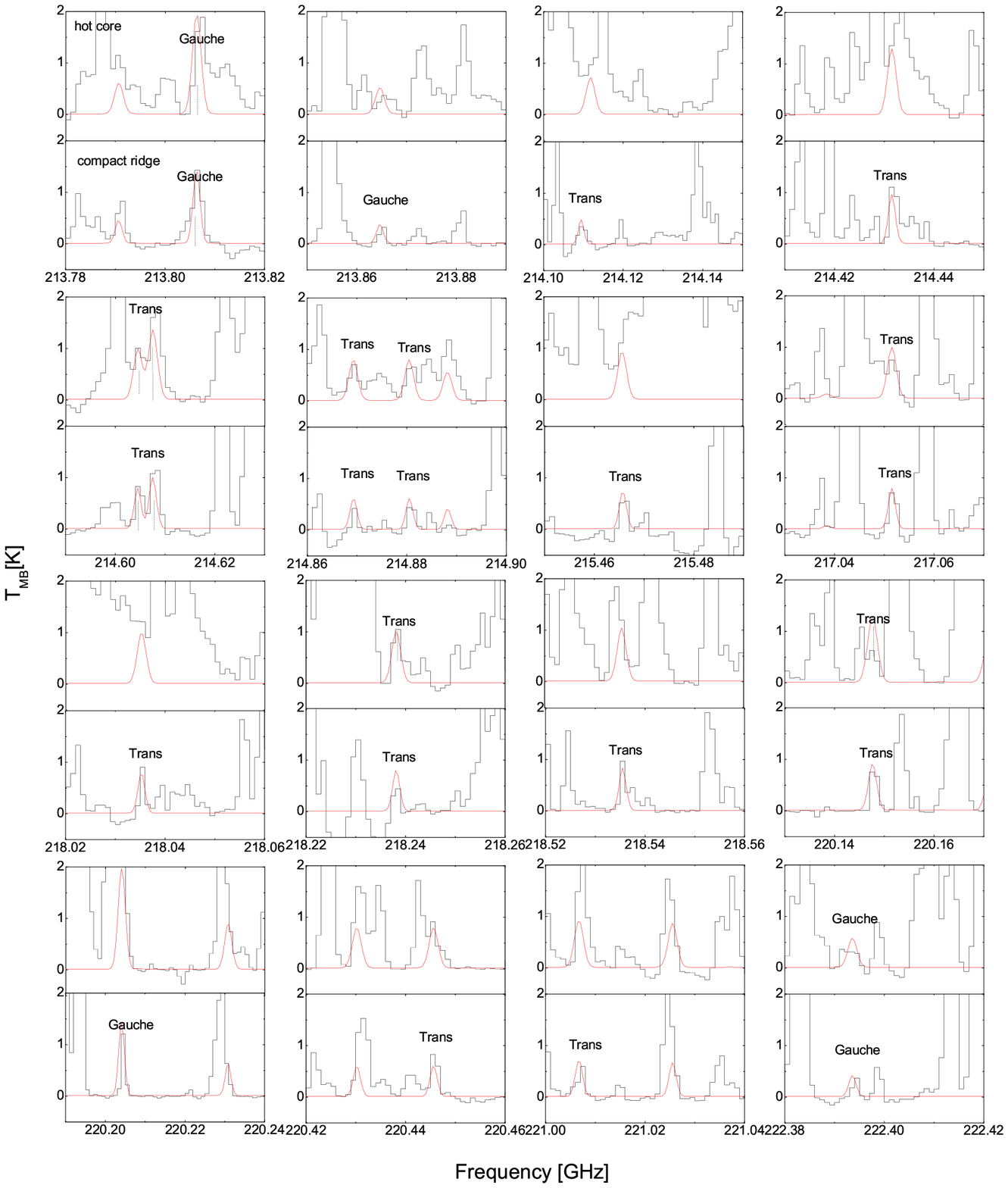}}}
\caption{Unblended or slightly blended transitions of the $trans$ and $gauche$ conformers of EF (C$_{2}$H$_{5}$OCHO) from transitions 213.807 GHz to 246.129 GHz toward the hot core-SW and/or compact ridge. The black curve is the observed spectra, and the red line indicates the simulated LTE spectra.
}
\label{1}
\setcounter{figure}{\value{1}}
\end{figure*}

\clearpage
\begin{figure*}
\setcounter{1}{\value{figure}}
\setcounter{figure}{0}
\centerline{\resizebox{1.0\hsize}{!}{\includegraphics{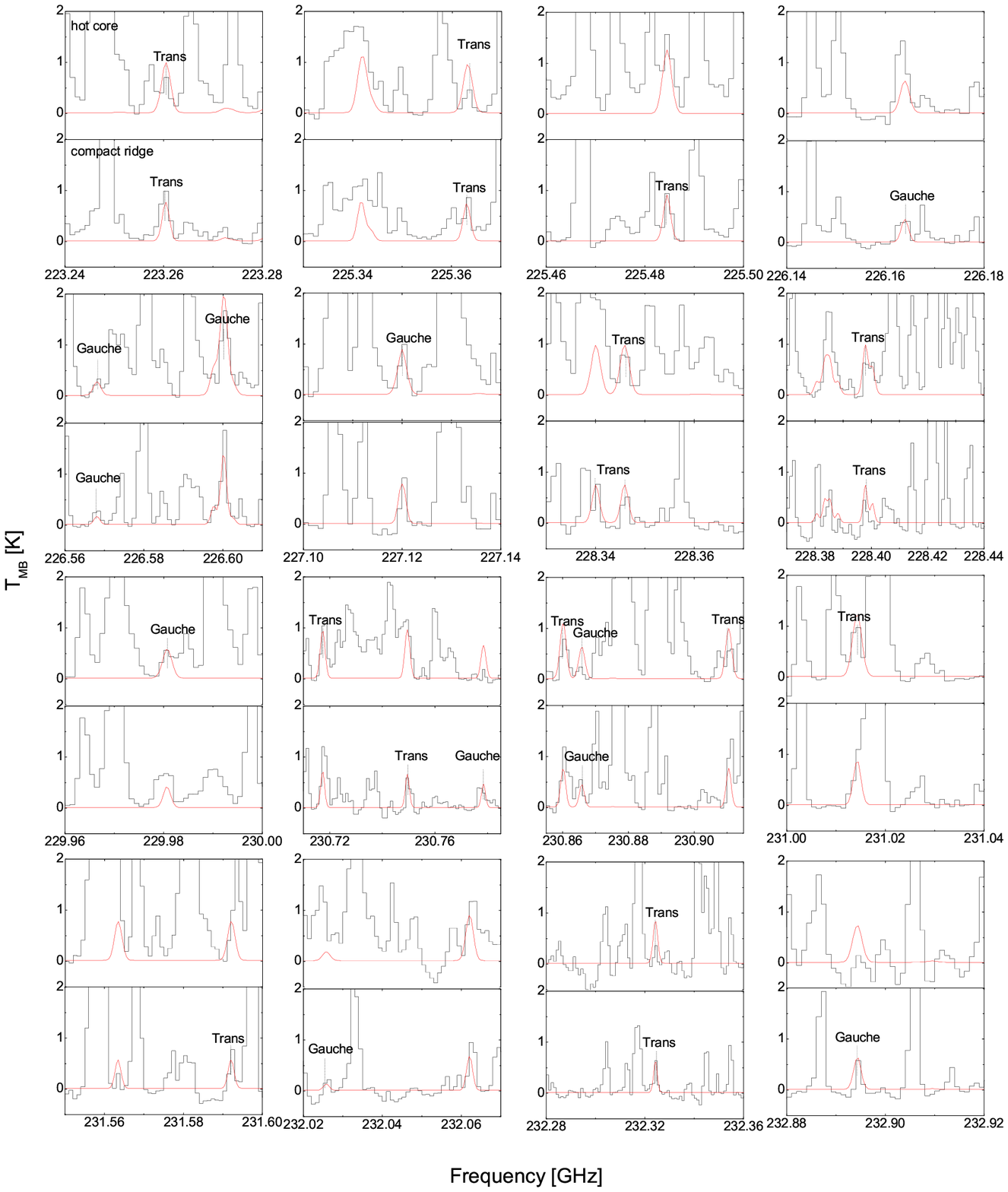}}}
\caption{Continued.
}
\label{f:survey_}
\end{figure*}

\clearpage
\begin{figure*}
\setcounter{1}{\value{figure}}
\setcounter{figure}{0}
\centerline{\resizebox{1.0\hsize}{!}{\includegraphics{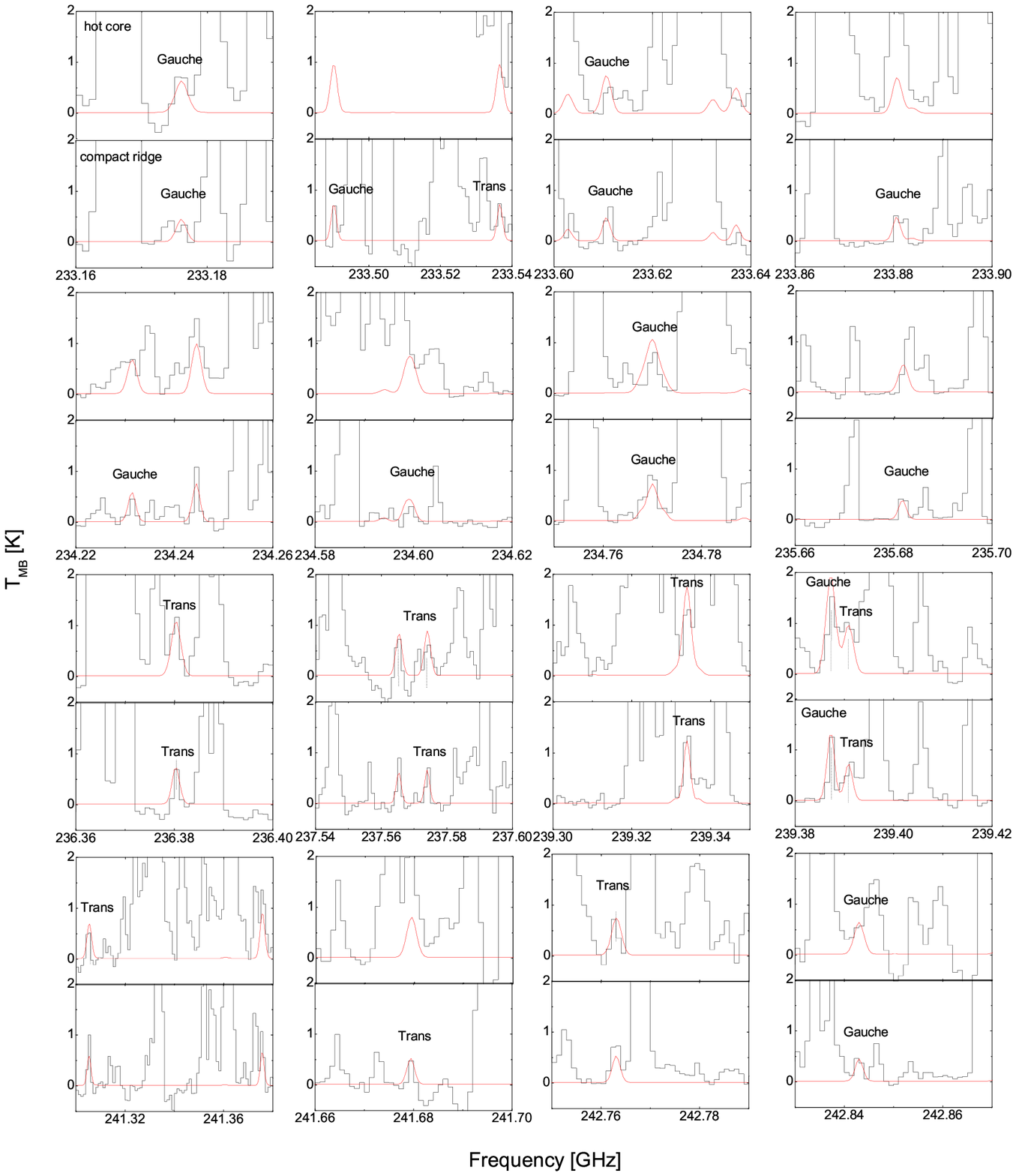}}}
\caption{Continued.
}
\label{f:survey_}
\end{figure*}

\clearpage
\begin{figure*}
\setcounter{1}{\value{figure}}
\setcounter{figure}{0}
\centerline{\resizebox{1.0\hsize}{!}{\includegraphics{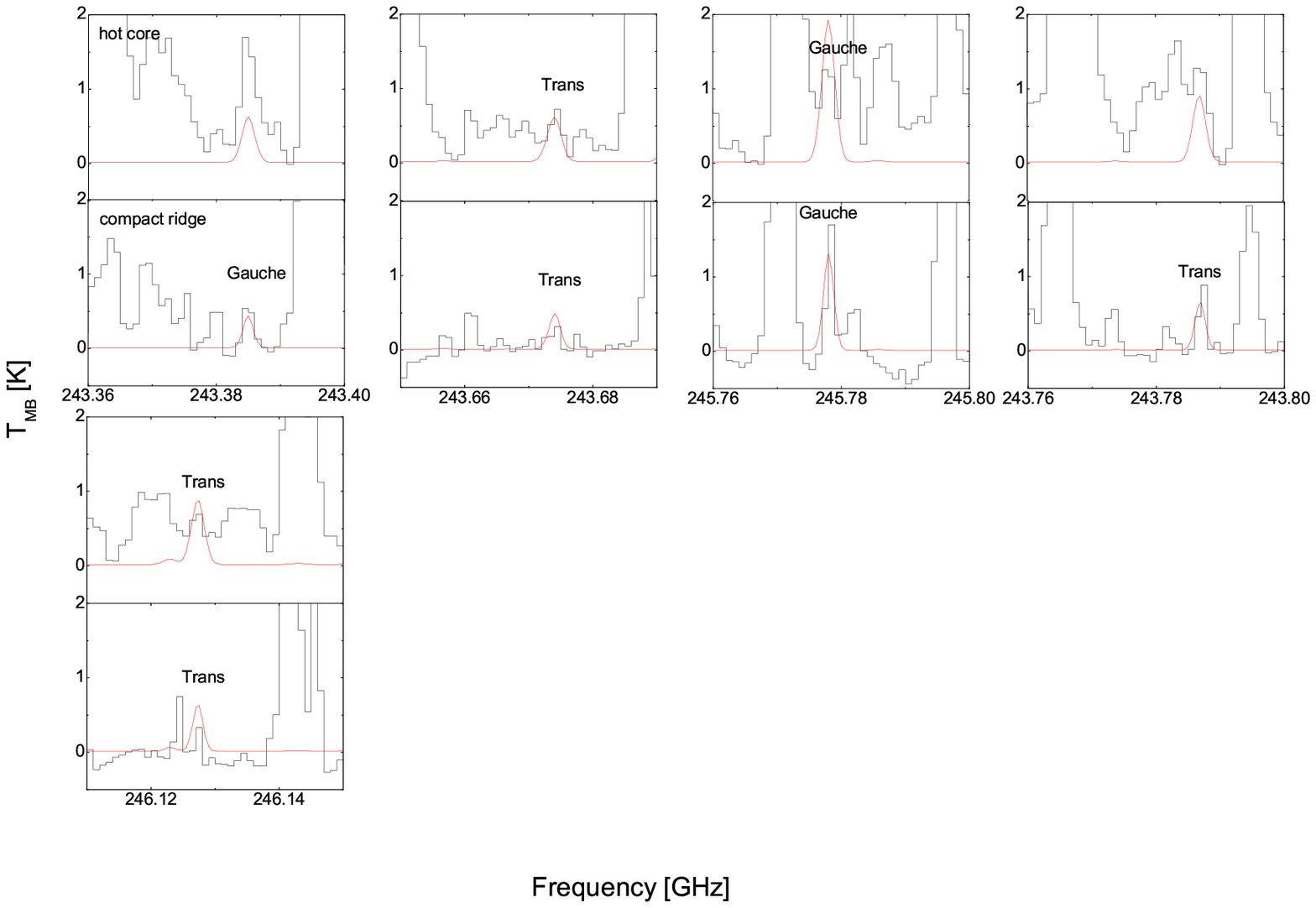}}}
\caption{Continued.
}
\label{f:survey_}
\end{figure*}

\clearpage
\begin{figure*}
\newcounter{2}
\setcounter{2}{\value{figure}}
\setcounter{figure}{1}
\centerline{\resizebox{0.9\hsize}{!}{\includegraphics{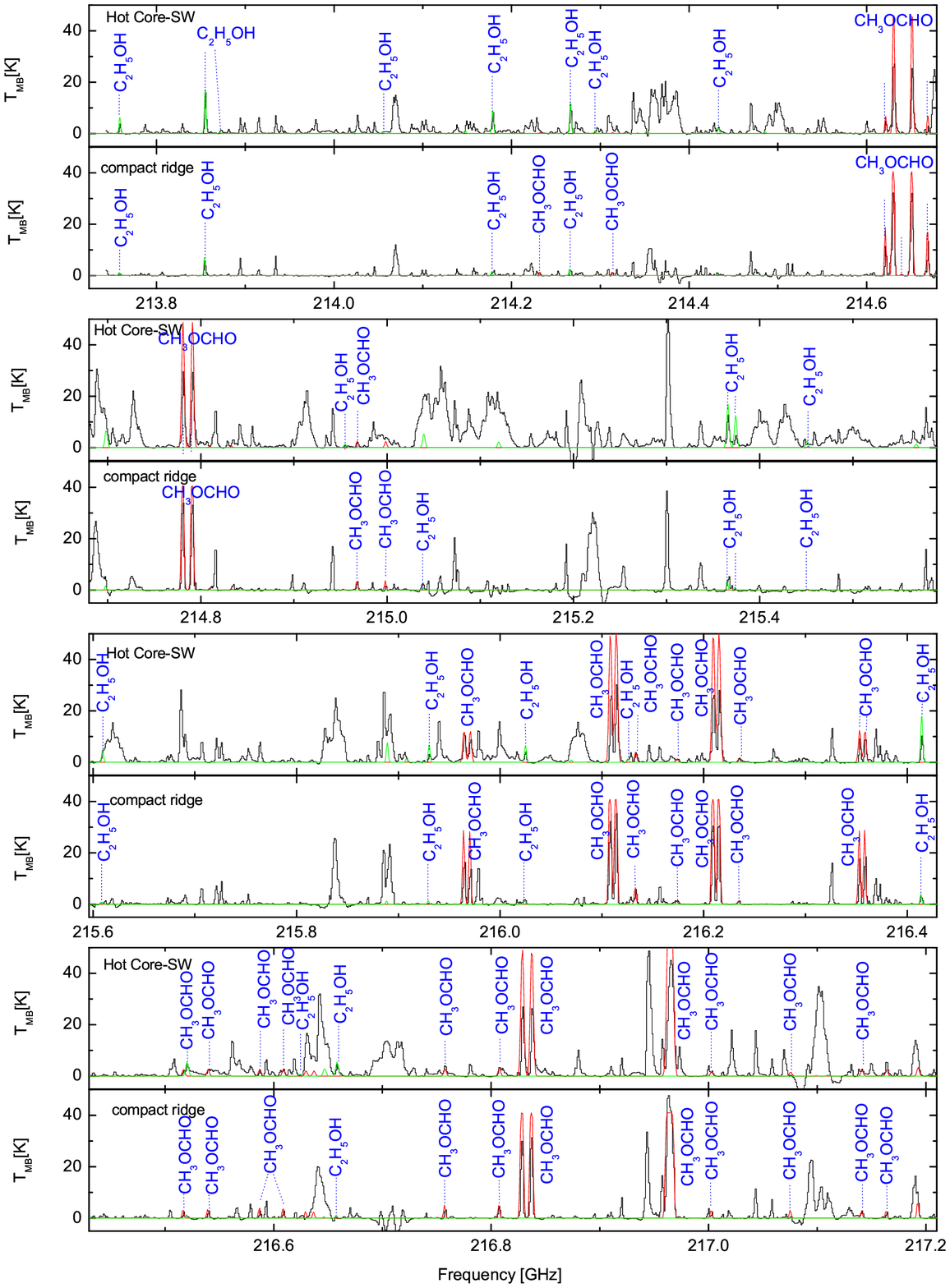}}}
\caption{Spectra of ET and MF detected in the Orion KL from 213 GHz to 217 GHz. Each panel consists of two plots: the top and bottom plots show the spectrum from the hot core and the compact ridge, respectively. The black curve is the observed spectra, and red curve (MF) and green curve (ET) indicate the simulated LTE spectra.
}
\label{2_}
\end{figure*}

\begin{figure*}
\setcounter{2}{\value{figure}}
\setcounter{figure}{1}
\centerline{\resizebox{1.0\hsize}{!}{\includegraphics{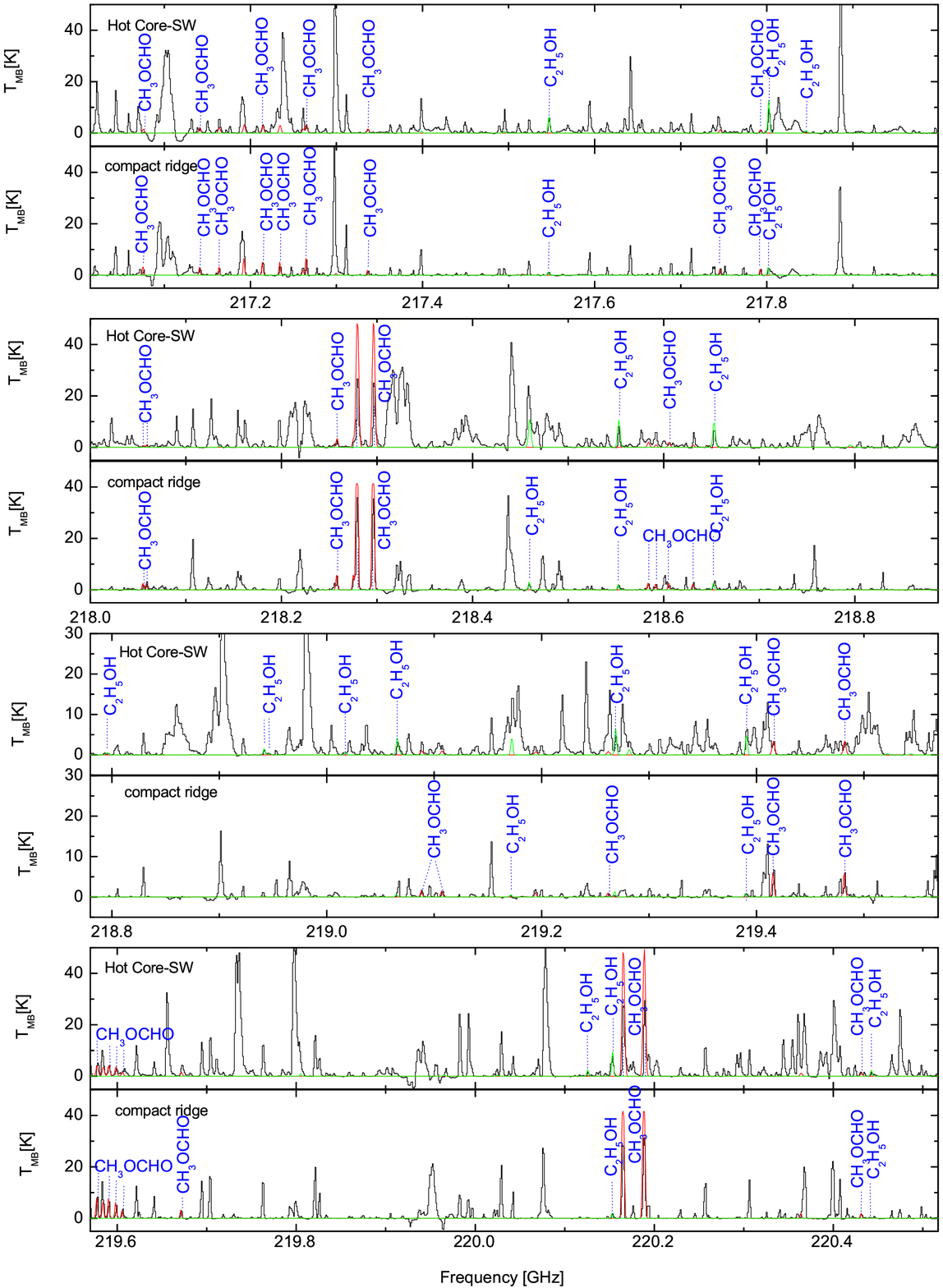}}}
\caption{
continued.
}
\label{f:survey_}
\end{figure*}

\begin{figure*}
\setcounter{2}{\value{figure}}
\setcounter{figure}{1}
\centerline{\resizebox{1.0\hsize}{!}{\includegraphics{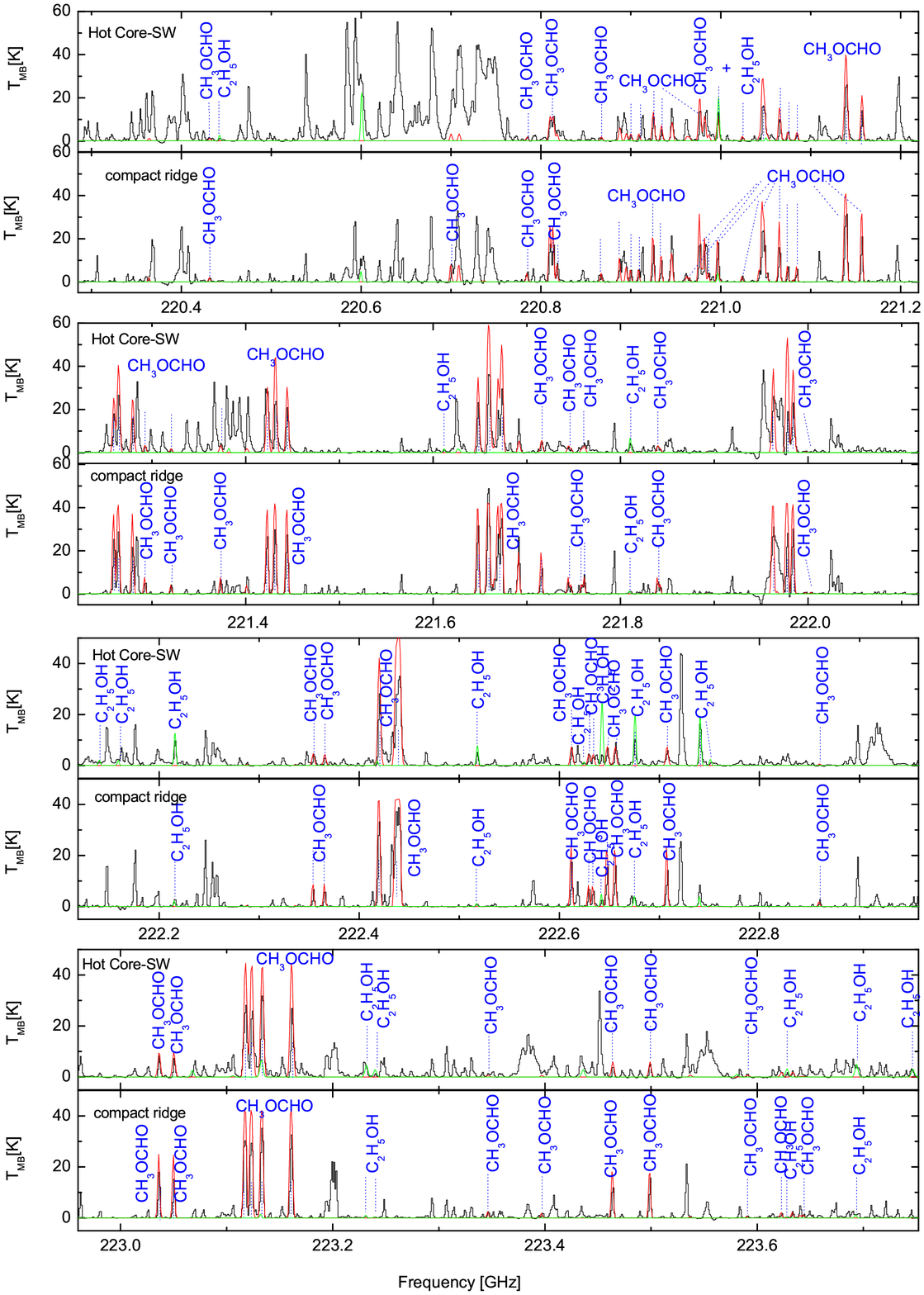}}}
\caption{
continued.
}
\label{f:survey_}
\end{figure*}

\begin{figure*}
\setcounter{2}{\value{figure}}
\setcounter{figure}{1}
\centerline{\resizebox{1.0\hsize}{!}{\includegraphics{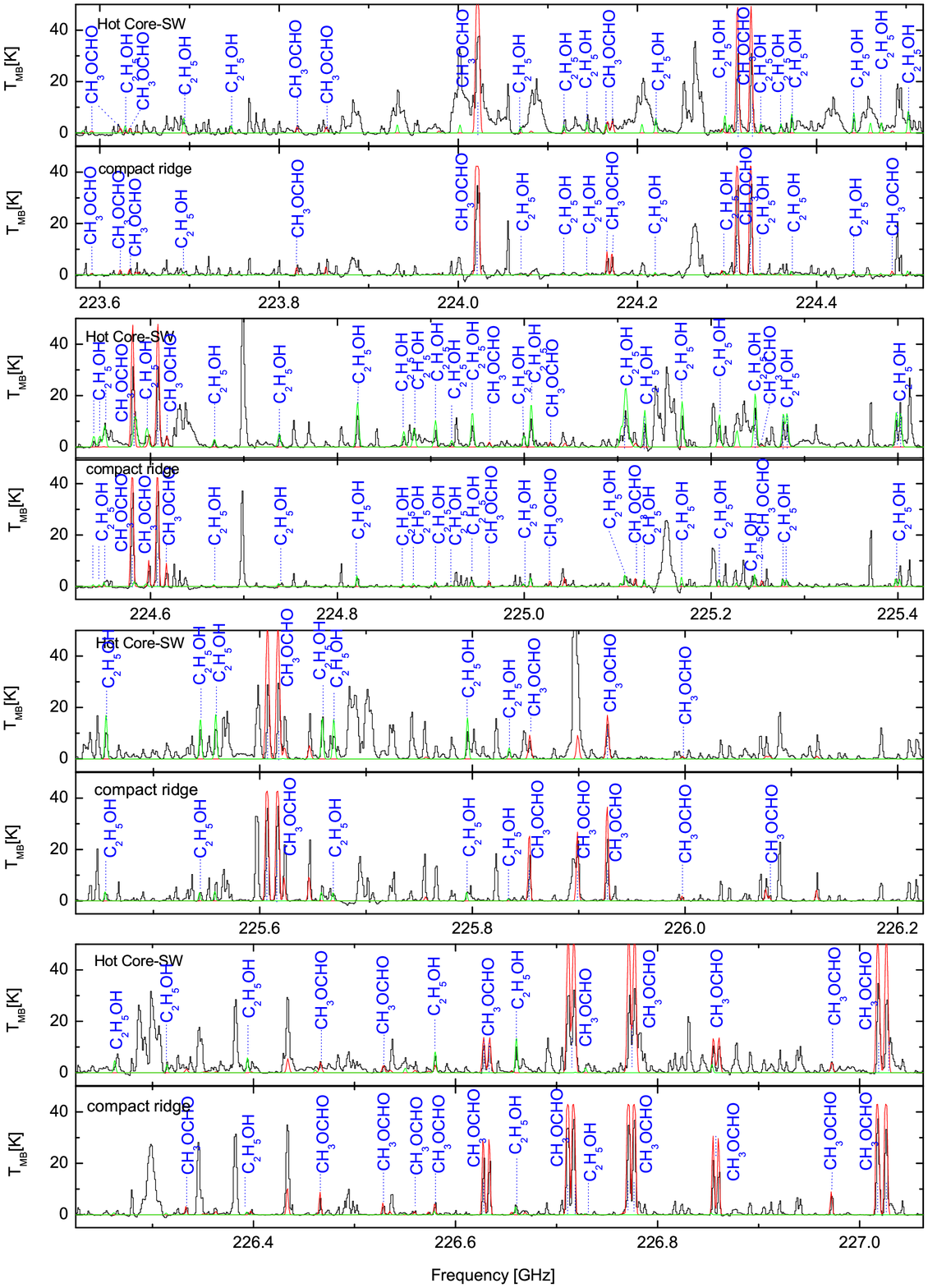}}}
\caption{
continued.
}
\label{f:survey_}
\end{figure*}

\begin{figure*}
\setcounter{2}{\value{figure}}
\setcounter{figure}{1}
\centerline{\resizebox{1.0\hsize}{!}{\includegraphics{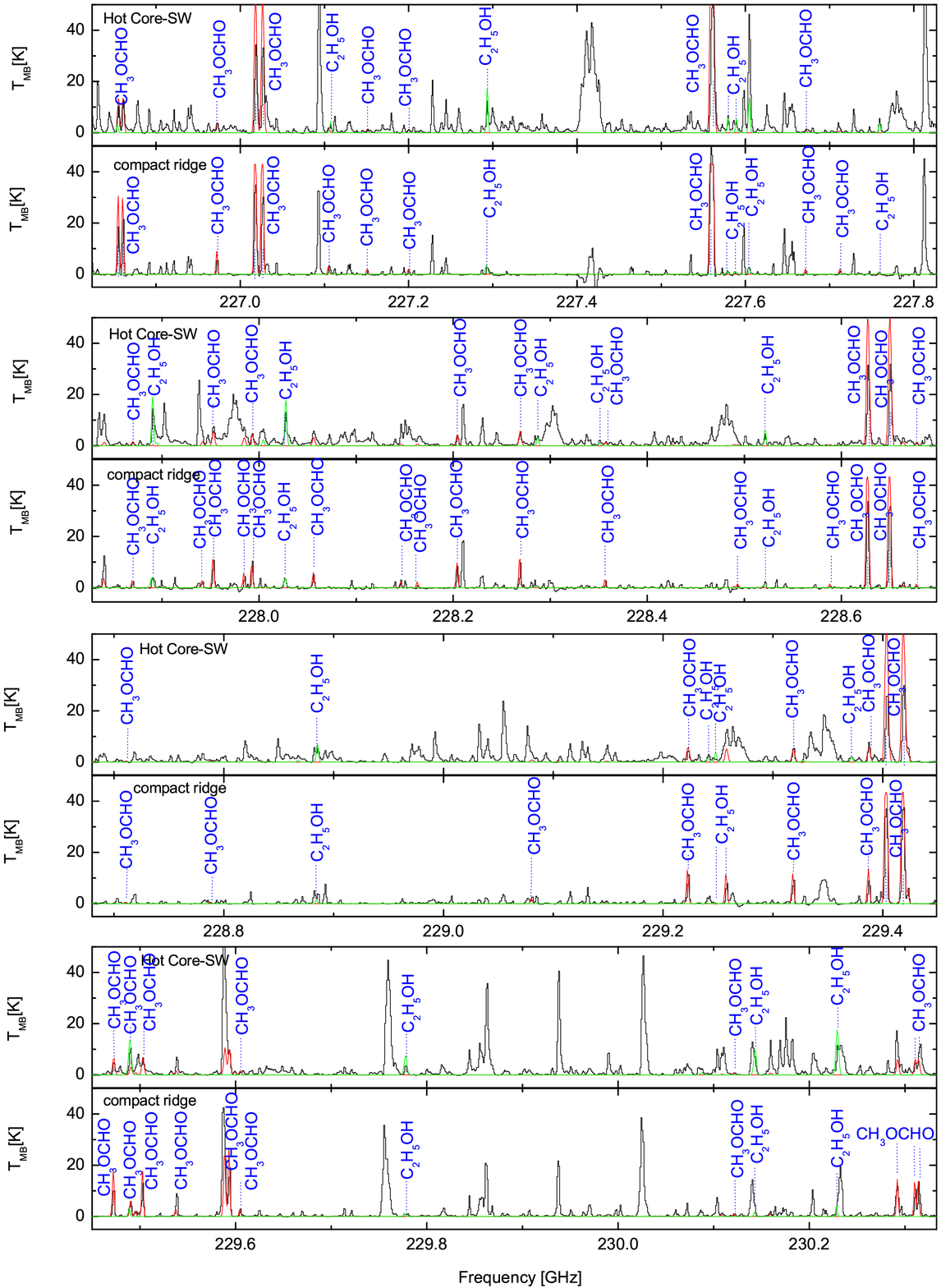}}}
\caption{
continued.
}
\label{f:survey_}
\end{figure*}

\begin{figure*}
\setcounter{2}{\value{figure}}
\setcounter{figure}{1}
\centerline{\resizebox{1.0\hsize}{!}{\includegraphics{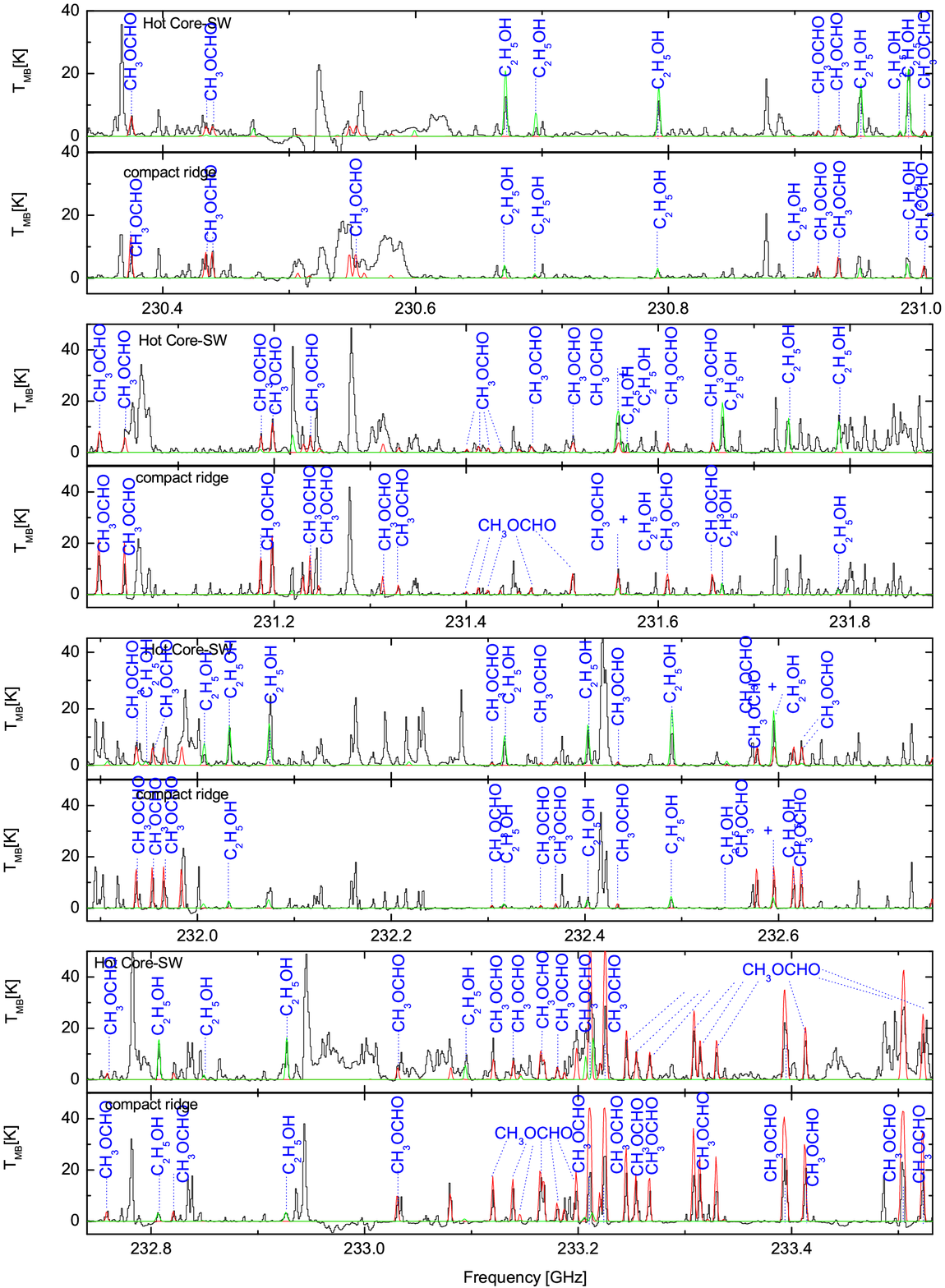}}}
\caption{
continued.
}
\label{f:survey_}
\end{figure*}

\begin{figure*}
\setcounter{2}{\value{figure}}
\setcounter{figure}{1}
\centerline{\resizebox{1.0\hsize}{!}{\includegraphics{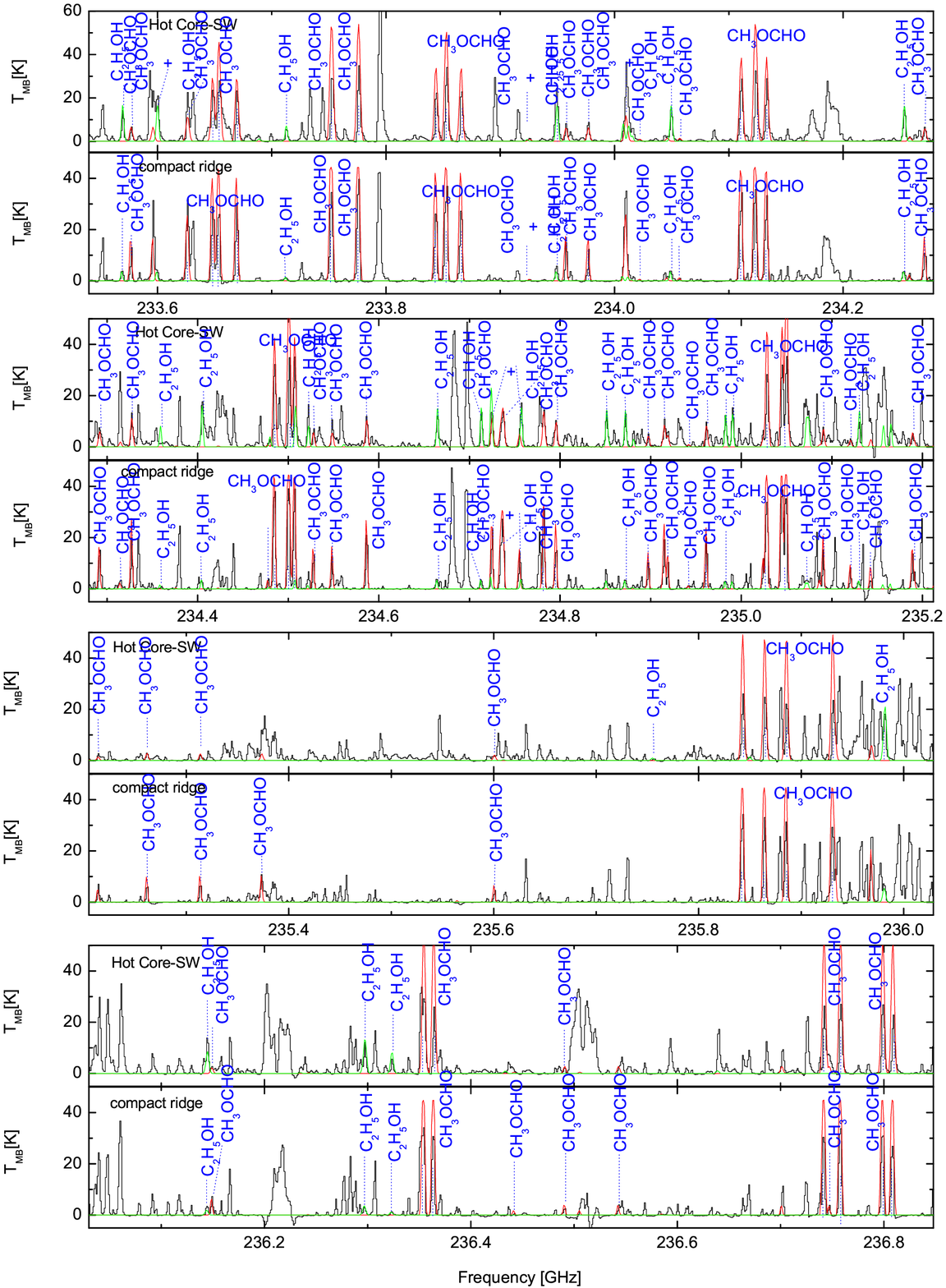}}}
\caption{
continued.
}
\label{f:survey_}
\end{figure*}

\begin{figure*}
\setcounter{2}{\value{figure}}
\setcounter{figure}{1}
\centerline{\resizebox{1.0\hsize}{!}{\includegraphics{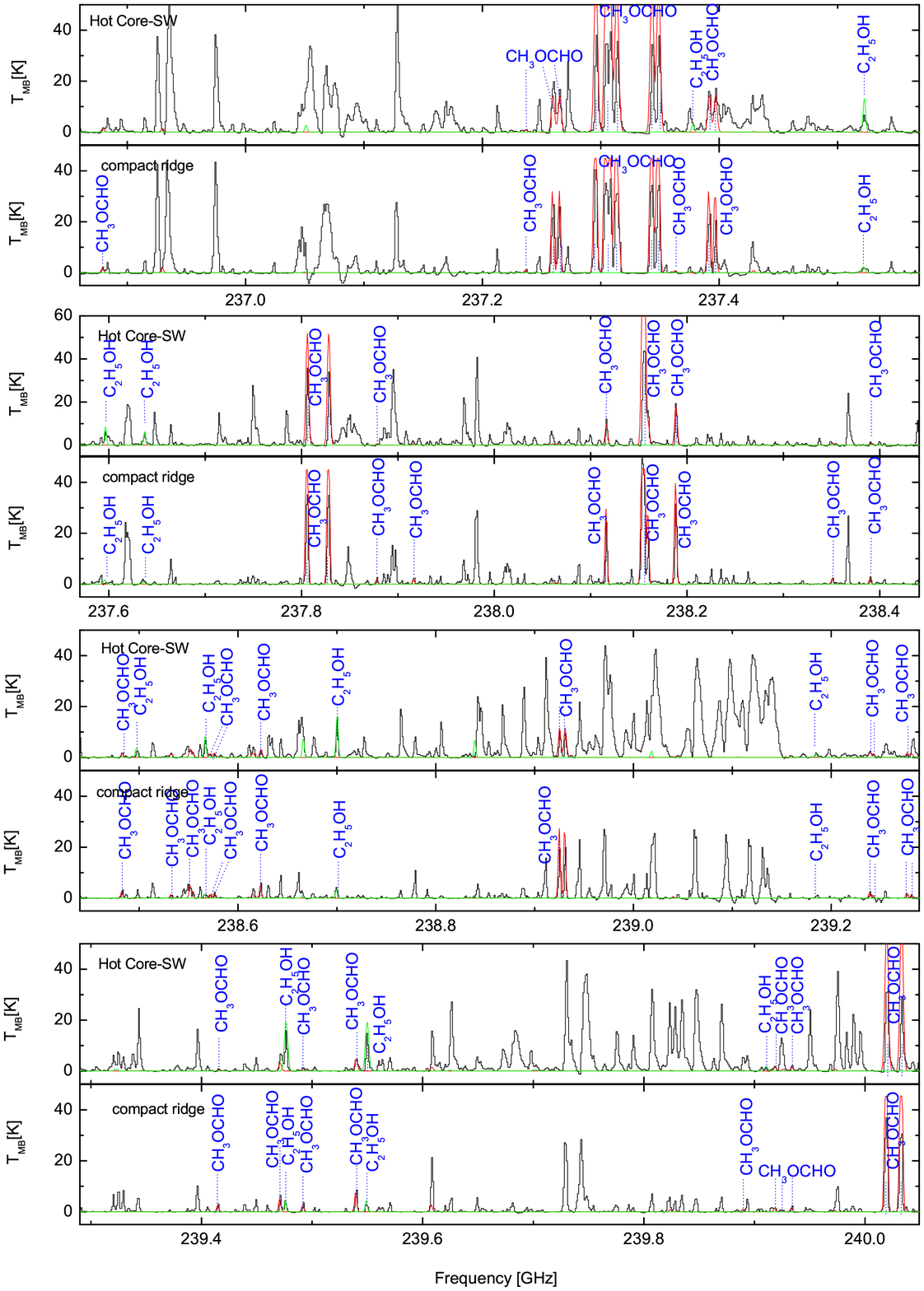}}}
\caption{
continued.
}
\label{f:survey_}
\end{figure*}

\begin{figure*}
\setcounter{2}{\value{figure}}
\setcounter{figure}{1}
\centerline{\resizebox{1.0\hsize}{!}{\includegraphics{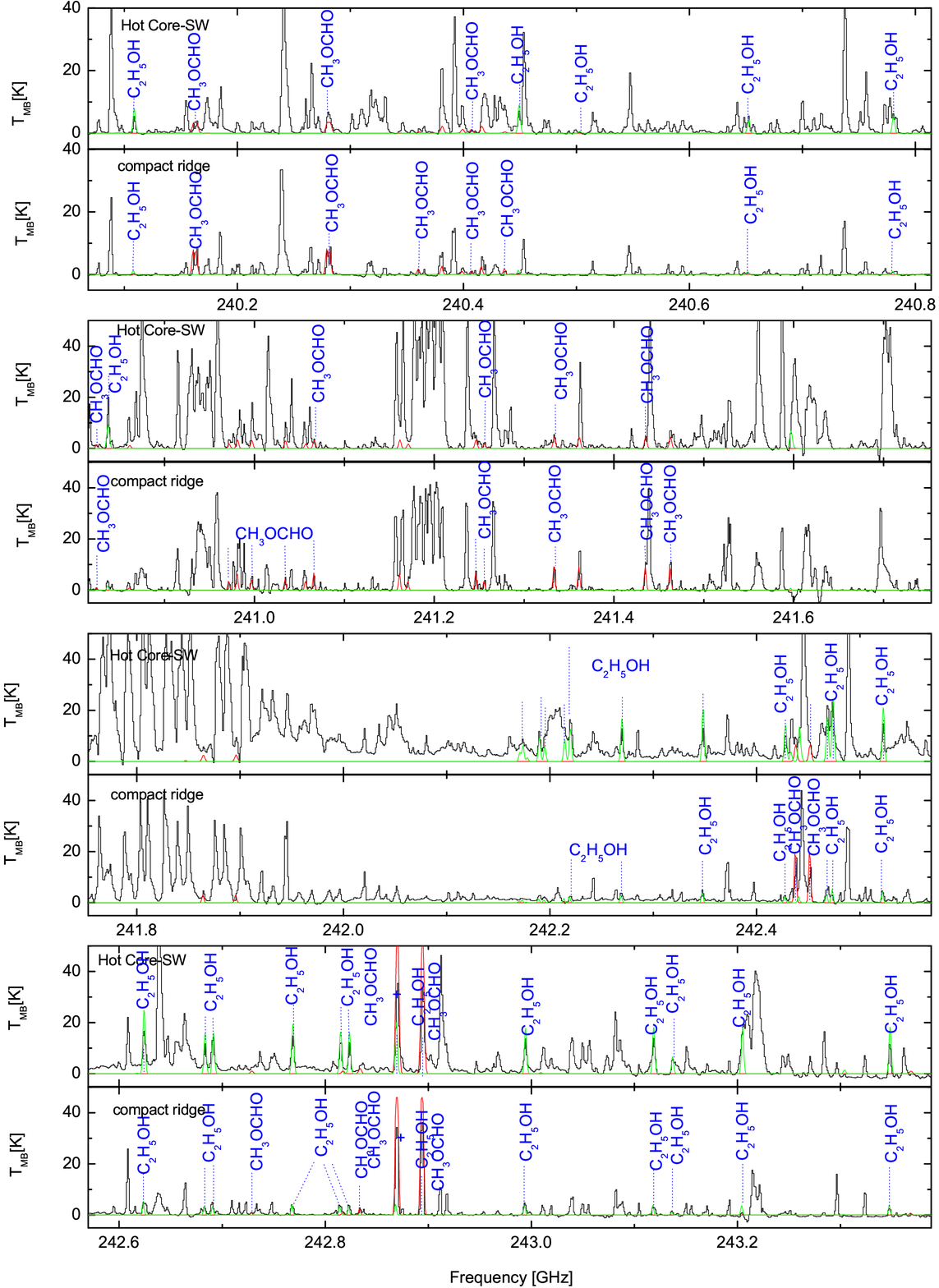}}}
\caption{
continued.
}
\label{f:survey_}
\end{figure*}

\begin{figure*}
\setcounter{2}{\value{figure}}
\setcounter{figure}{1}
\centerline{\resizebox{1.0\hsize}{!}{\includegraphics{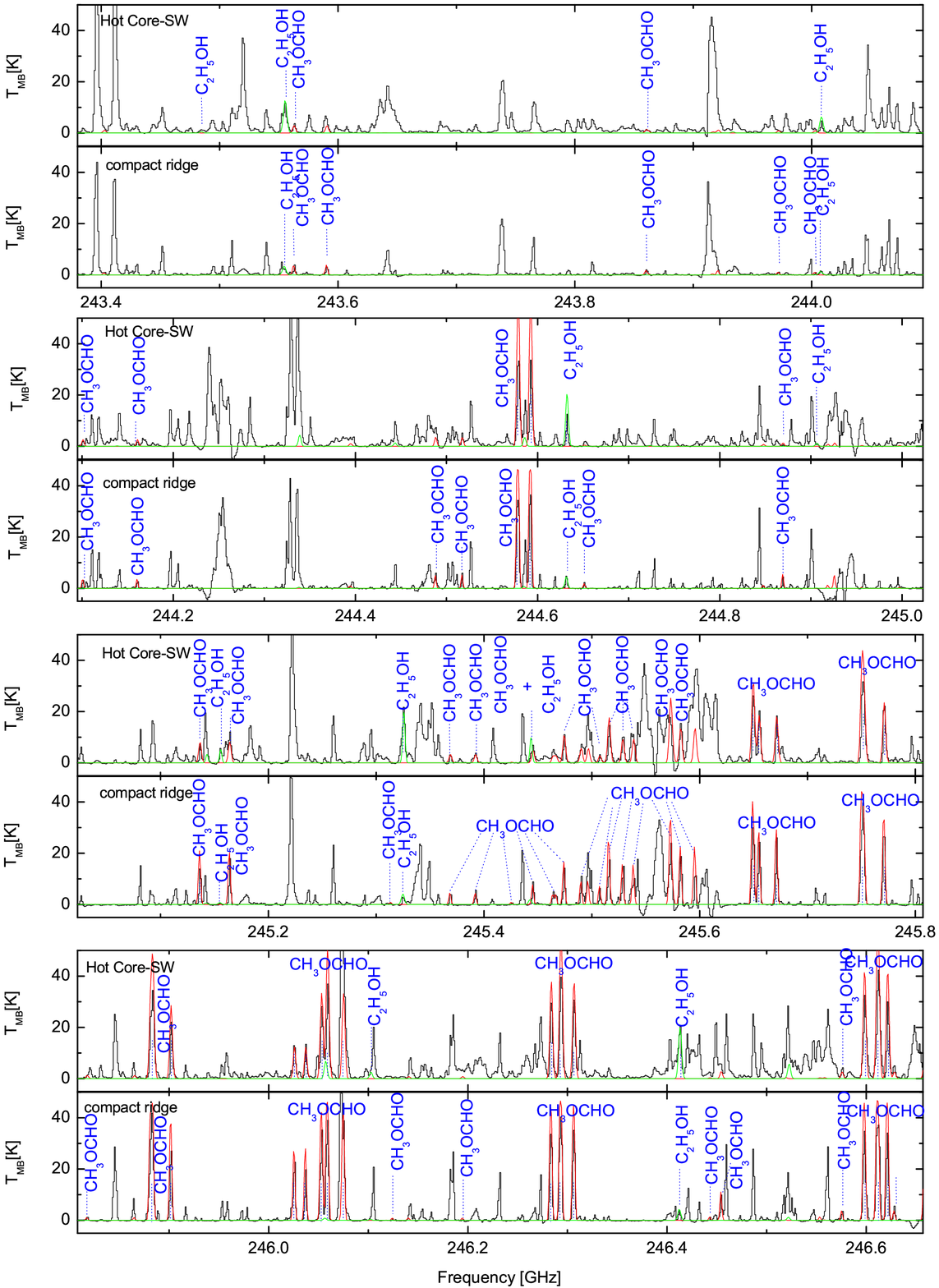}}}
\caption{
continued.
}
\label{f:survey_}
\end{figure*}

\begin{figure*}
\begin{center}
\centerline{\resizebox{1.0\hsize}{!}{\includegraphics{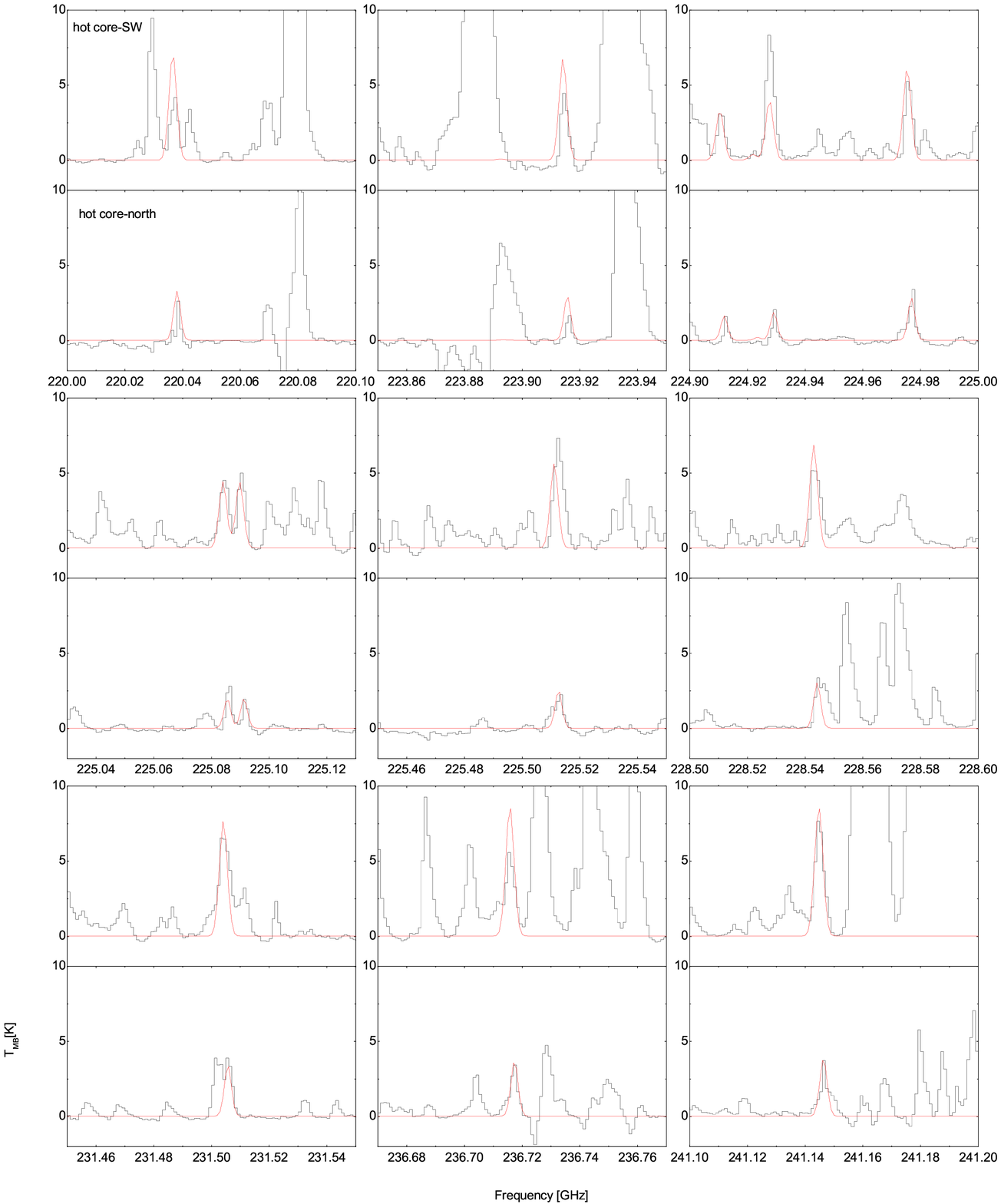}}}
\end{center}
\caption{ Spectra of twelve transitions of t-HCOOH toward the hot core-SW and hot core-N. The black curve is the observed spectra, and red curve indicates the simulated LTE spectra.}
   \label{Fig3}
\end{figure*}

\begin{figure*}
\begin{center}
\hspace{-0.69cm}
\includegraphics[width=130mm,height=150mm,angle=270]{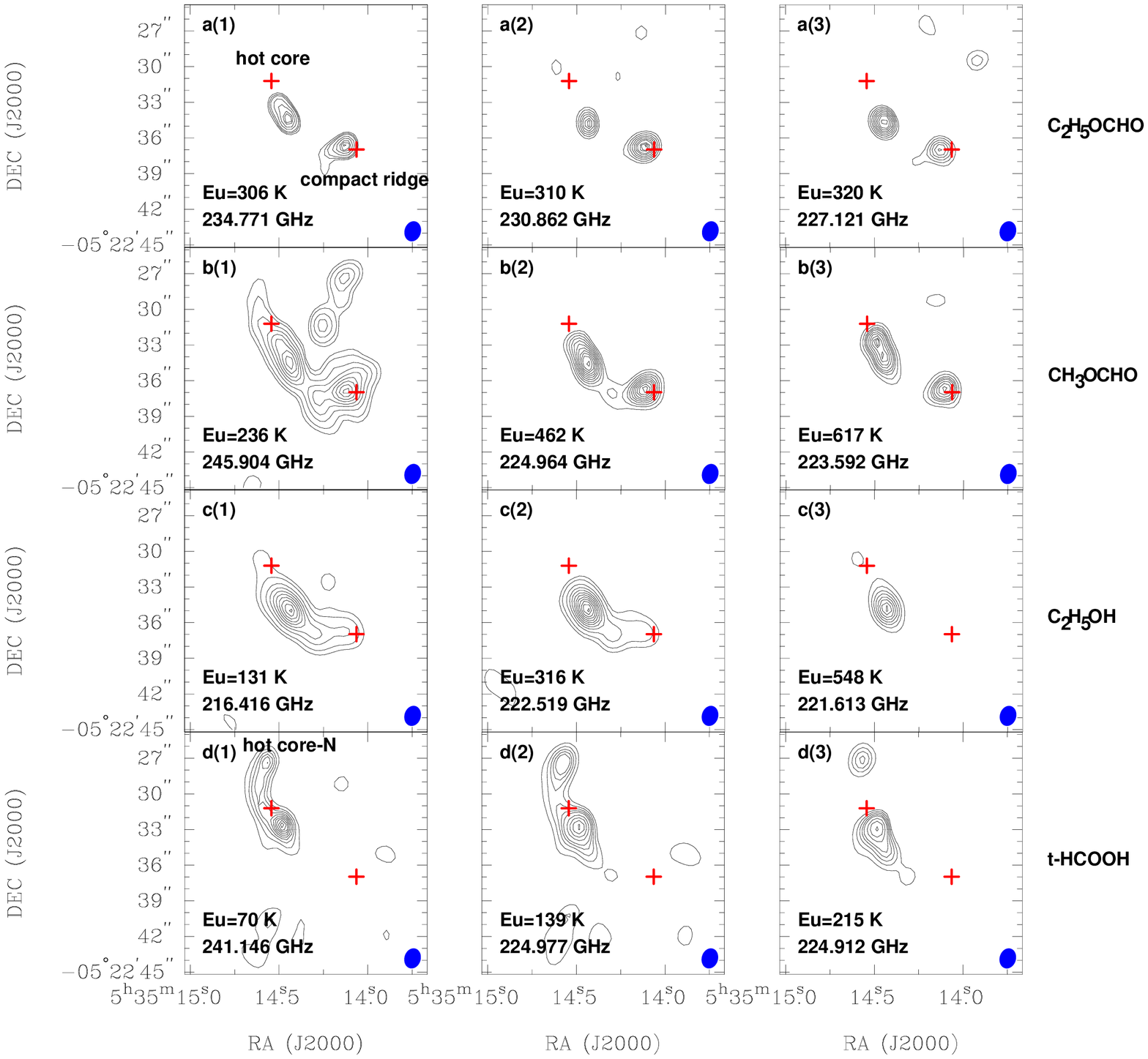}
\end{center}
\caption{ The velocity integrated intensity maps of EF (panel a), MF (panel b), ET (panel c) and t-HCOOH (panel d) at different upper level energies. The upper state energy and frequency are indicated on each plot. The synthesized beam is shown in the bottom right corner. The contour levels are: a(1) (5, 6, 7, 9, 10, 11, 12, 12.8) $\times$ 0.02 Jy~beam$^{-1}$ km~s$^{-1}$, a(2) (5, 7, 9, 11, 13, 15, 16.5, 19, 20, 21) $\times$ 0.007 Jy~beam$^{-1}$ km~s$^{-1}$, a(3) (5, 7, 9, 11, 13, 15, 17, 19, 22) $\times$ 0.008 Jy~beam$^{-1}$ km~s$^{-1}$, b(1) (5, 10, 20, 30, 40, 50, 60, 70, 78) $\times$ 0.095 Jy~beam$^{-1}$ km~s$^{-1}$, b(2) (5, 8, 11, 14, 17, 20, 23, 26, 29, 32, 36) $\times$ 0.014 Jy~beam$^{-1}$ km~s$^{-1}$, b(3) (5, 8, 11, 13, 15, 18, 21, 24, 27) $\times$ 0.01 Jy~beam$^{-1}$ km~s$^{-1}$, c(1) (5, 10, 20, 30, 40, 50, 60, 70, 80, 87) $\times$ 0.05 Jy~beam$^{-1}$ km~s$^{-1}$, c(2) (5, 10, 20, 30, 40, 50, 60, 70, 80, 90, 100, 108) $\times$ 0.018 Jy~beam$^{-1}$ km~s$^{-1}$, c(3) (5, 9, 13, 17, 21, 25, 29, 32.5) $\times$ 0.018 Jy~beam$^{-1}$ km~s$^{-1}$, d(1) (5, 10, 15, 20, 25, 30, 35, 40, 45, 50) $\times$ 0.045 Jy~beam$^{-1}$ km~s$^{-1}$, d(2) (5, 10, 15, 20, 25, 35, 45, 55, 65) $\times$ 0.028 Jy~beam$^{-1}$ km~s$^{-1}$, and d(3)(5, 10, 15, 20, 25, 35, 45, 57) $\times$ 0.02 Jy~beam$^{-1}$ km~s$^{-1}$. The cross symbols mark the continuum peak positions of hot core and compact ridge. The coordinates of two continuum peaks are presented in Section 3.2.}
   \label{Fig4}
\end{figure*}

\begin{figure*}
\begin{center}
\centerline{\resizebox{1.0\hsize}{!}{\includegraphics{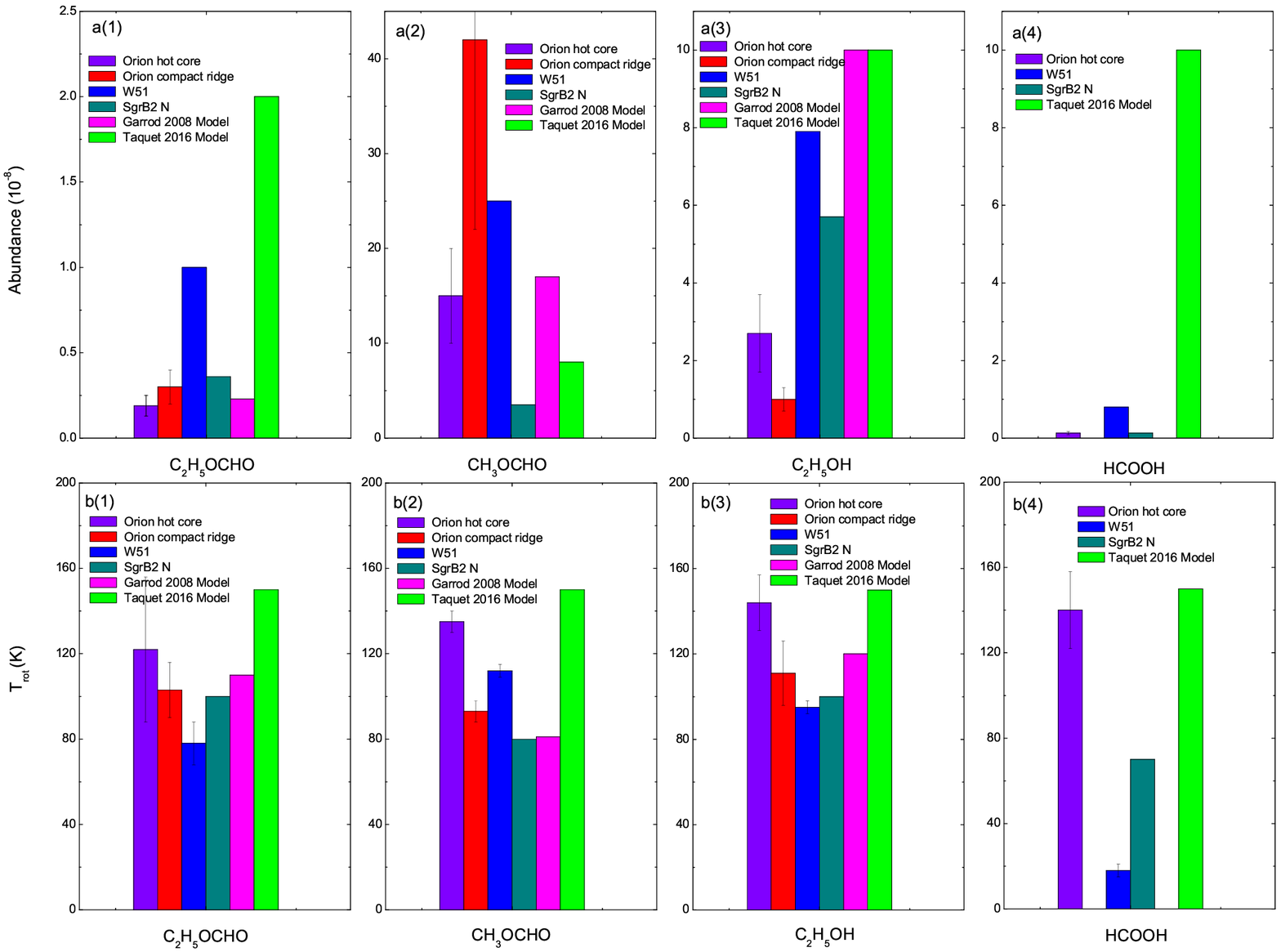}}}
\end{center}
\caption{The abundances (top panels) and rotational temperatures (bottom panels) of each molecule (EF, MF, ET and HCOOH) derived in different regions (Orion KL, W51 and Sgr B2(N)), and the theoretical values from the grain-surface model by \citet{Garrod08} and gas-phase model by \citet{Taquet16}. The theoretical abundances of around 2$\times$10$^{-8}$ for EF, 8$\times$10$^{-8}$ for MF, 1$\times$10$^{-7}$ for ET and a few 10$^{-7}$ for HCOOH, in a reasonable timescale for hot core sources of 5 $\times$ 10$^{4}$ yr, from \citet{Taquet16} are the values obtained by considering PT reactions. For the estimated abundances and temperatures of EF, MF and ET in grain-surface model, the values are from \citet{Belloche09} (in Basic model, see their Table~15), which uses \citet{Garrod08} chemical network.}
   \label{Fig3}
\end{figure*}

 \clearpage

\appendix

   \label{sec-part}
\begin{deluxetable}{l l c c c c c c c c c c c c c c c c c c c c c c c c c c }
\tabletypesize{\tiny}
\tablenum{0}
\setcounter{table}{0}
\renewcommand\thetable{A.\arabic{table}}
\rotate
\tablewidth{-30pt}
\tablecaption{ Selected unblended transitions of the $Trans$ and $Gauche$ conformers of EF observed with the ALMA. }
\tablehead{\colhead{Rest Frequency} & \colhead{Transition} &\colhead{$S_{ij}\mu^{2}$} & \colhead{E$_{\rm u }$}
           &\multicolumn{4}{c}{hot core-SW}& &\multicolumn{4}{c}{compact ridge}& \colhead{$\sigma$}\\
           \cline{5-8}  \cline{10-13} \\
                      & & & & \colhead{$v_{\rm LSR}$}& \colhead{$\Delta V $} &\colhead{$I_{\rm p}$} & \colhead{Detection}& & \colhead{$v_{\rm LSR}$} & \colhead{$\Delta V $} &\colhead{$I_{\rm p}$} &\colhead{Detection} \\
                      (GHz) & &\colhead{(Debye$^{2})$} & \colhead{(K)}  & \colhead{(km s$^{-1}$)} & \colhead{(km s$^{-1}$)} &\colhead{(K)} & \colhead{Level} & & \colhead{(km s$^{-1}$)} & \colhead{(km s$^{-1}$)} & \colhead{(K)}  & \colhead{Level}&\colhead{(K)}}

\startdata
 \multicolumn{14}{c}{$Trans$ conformer}\\
 \hline
214.11074 & $39(13, 26)-38(13, 25)$ & 118.64 & 327 & \multicolumn{4}{c}{B} & & 7.2$\pm$0.1 & 1.7$\pm$0.3 & 0.42$\pm$0.07 & 7$\sigma$  & 0.06 \\
214.43282 & $39(9, 31)-38(9, 30)$ & 126.36 & 264 & \multicolumn{4}{c}{\multirow{2}{*}{B}} & & \multirow{2}{*}{7.0$\pm$0.1} & \multirow{2}{*}{2.0$\pm$0.1} & \multirow{2}{*}{1.09$\pm$0.05} & \multirow{2}{*}{23$\sigma$}  & \multirow{2}{*}{0.05}\\
214.43284 & $39(9, 30)-38(9, 29)$ & 126.36 & 264 &  \\
214.60560 & $39(4, 36)-38(4, 35)$ & 131.98 & 218 & \multicolumn{4}{c}{B} & & 7.0$\pm$0.1 & 2.0$\pm$0.1 & 0.65$\pm$0.05 & 11$\sigma$ & 0.06\\
214.60858 & $39(8, 32)-38(8, 31)$ & 127.85 & 251 & \multirow{2}{*}{6.5$\pm$0.1} & \multirow{2}{*}{4.2$\pm$0.1} & \multirow{2}{*}{1.40$\pm$0.04} &  \multirow{2}{*}{25$\sigma$} & & \multirow{2}{*}{6.9$\pm$0.1} & \multirow{2}{*}{2.5$\pm$0.1} & \multirow{2}{*}{1.15$\pm$0.05} & \multirow{2}{*}{20$\sigma$} &  \multirow{2}{*}{0.06}\\
214.60911 & $39(8, 31)-38(8, 30)$ & 127.85 & 251 & \\
214.87062 & $39(7, 33)-38(7, 32)$ & 129.18 & 241 & 7.3$\pm$0.2 & 4.0$\pm$0.4 & 0.78$\pm$0.05 & 14$\sigma$ & & 7.4$\pm$0.1 & 1.7$\pm$0.3 & 0.59$\pm$0.06 & 10$\sigma$ & 0.06\\
\enddata
\tablecomments{Abbreviations in the "Detection" column are B: blended; PB: partial blend. }
\end{deluxetable}

\begin{deluxetable}{c l c c c c c c c c c}
\tablenum{1}
\setcounter{table}{1}
\renewcommand\thetable{A.\arabic{table}}
\tablewidth{0pt}
\tablecaption{Selected unblended transitions of MF observed with the ALMA. }
\tablehead{\colhead{Rest Frequency} & \colhead{Transition} &\colhead{$S_{ij}\mu^{2}$} & \colhead{E$_{\rm u }$}
           &\multicolumn{2}{c}{Detection} \\
           \cline{5-6} \\
                      (GHz) & &\colhead{(Debye$^{2})$} & \colhead{(K)}&\colhead{hot core-SW} &\colhead{compact ridge} }

\startdata
214.23362 & $13(4, 9)-13(1, 12)E$ & 0.13 & 65 & B & $\surd$ \\
214.31522 & $13(4, 9)-13(1, 12)A$ & 0.13 & 65 & B & $\surd$ \\
214.62285 & $10(4, 7)-9(3, 6)E$ & 1.71 & 43 & $\surd$ & $\surd$  \\
214.63169 & $17(5, 12)-16(5, 11)E$ & 41.28 & 108 & $\surd$ & $\surd$ \\
214.65259 & $17(5, 12)-16(5, 11)A$ & 41.28 & 108 & $\surd$ & $\surd$ \\
214.67036 & $10(4, 7)-9(3, 6)A$ & 1.75 & 43 & $\surd$ & $\surd$ \\
214.78236 & $18(3, 16)-17(3, 15)E$ & 46.05 & 106 & $\surd$ & $\surd$ \\
214.79249 & $18(3, 16)-17(3, 15)A$ & 45.98 & 106 & $\surd$ & $\surd$ \\
\enddata
\end{deluxetable}

\begin{deluxetable}{c l c c c c c c c c c}
\tablenum{2}
\setcounter{table}{2}
\renewcommand\thetable{A.\arabic{table}}
\tablewidth{0pt}
\tablecaption{ Transitions of the $Trans$ and $Gauche$ conformers of ET observed with the ALMA. }
\tablehead{\colhead{Rest Frequency} & \colhead{Transition} &\colhead{$S_{ij}\mu^{2}$} & \colhead{E$_{\rm u }$}
           &\multicolumn{2}{c}{Detection} \\
           \cline{5-6} \\
                      (GHz) & &\colhead{(Debye$^{2})$} & \colhead{(K)}&\colhead{hot core-SW} &\colhead{compact ridge} }

\startdata
 \multicolumn{6}{c}{$Trans$ conformer}\\
 \hline
213.85626 & $13(0, 13)-12(1, 12)$ & 12.83 & 74 & $\surd$ & $\surd$  \\
214.18012 & $22(5, 17)-22(4, 18)$ & 23.10 & 245 & $\surd$ & $\surd$  \\
214.26774 & $9(2, 8)- 8(1, 7)$ & 6.53 & 43 & $\surd$ & $\surd$ \\
214.95629 & $18(11, 7)-19(10, 10)$ & 2.02 & 295 & $\surd$ & ND \\
215.37564 & $21(2, 20)-21(1, 21)$ & 22.88 & 196 & $\surd$ & $\surd$ \\
215.45224 & $26(4, 23)-25(5, 20)$ & 9.50 & 316 & $\surd$ & $\surd$  \\
215.61072 & $20(3, 17)-19(4, 16)$ & 7.27 & 191 & B & $\surd$  \\
\enddata

\end{deluxetable}

\begin{deluxetable}{c l c c c c c c c c c}
\tablenum{3}
\setcounter{table}{3}
\renewcommand\thetable{A.\arabic{table}}
\tablewidth{0pt}
\tablecaption{ Unblended lines of the formic acid (t-HCOOH) toward the hot core-SW and hot core-north. }
\tablehead{\colhead{Rest Frequency} & \colhead{Transition} &\colhead{$S_{ij}\mu^{2}$} & \colhead{E$_{\rm u }$}
           &\multicolumn{2}{c}{Detection} \\
           \cline{5-6} \\
                      (GHz) & &\colhead{(Debye$^{2})$} & \colhead{(K)}&\colhead{hot core-SW} &\colhead{hot core-north} }

\startdata
220.03797 & $10(0, 10)-9(0, 9)$ & 20.14 & 59 & $\surd$ & $\surd$\\
223.91557 & $10(2, 9)-9(2, 8)$ & 19.39 & 72 & $\surd$ & $\surd$\\
224.91183 & $10(7, 3)-9(7, 2)$ & 10.30 & 215 & $\surd$ & $\surd$\\
224.92908 & $10(6, 5)-9(6, 4)$ & 12.93 & 174 & B & $\surd$\\
224.97676 & $10(5, 6)-9(5, 5)$ & 15.15 & 139 & $\surd$ & $\surd$\\
225.08544 & $10(4, 7)- 9(4, 6)$ & 16.97 & 110 & $\surd$ & $\surd$\\
225.09121 & $10(4, 6)- 9(4, 5)$ & 16.97 & 110 & $\surd$ & $\surd$\\
225.51255 & $10(3, 7)- 9(3, 6)$ & 18.38 & 88 & B & $\surd$\\
228.54408 & $10(2, 8)-9(2, 7)$ & 19.40 & 73 & $\surd$ & B\\
231.50561 & $10(1, 9)- 9(1, 8)$ & 19.99 & 64 & $\surd$ & B\\
236.71720 & $11(1, 11)-10(1, 10)$ & 22.03 & 72 & $\surd$ & $\surd$\\
241.14625 & $11(0, 11)-10(0, 10)$ & 22.15 & 70 & $\surd$ &$\surd$\\
\enddata

\end{deluxetable}

\begin{figure*}
\newcounter{6}
\setcounter{6}{\value{figure}}
\setcounter{figure}{0}
\renewcommand\thefigure{A.\arabic{figure}}
\centerline{\resizebox{1.0\hsize}{!}{\includegraphics{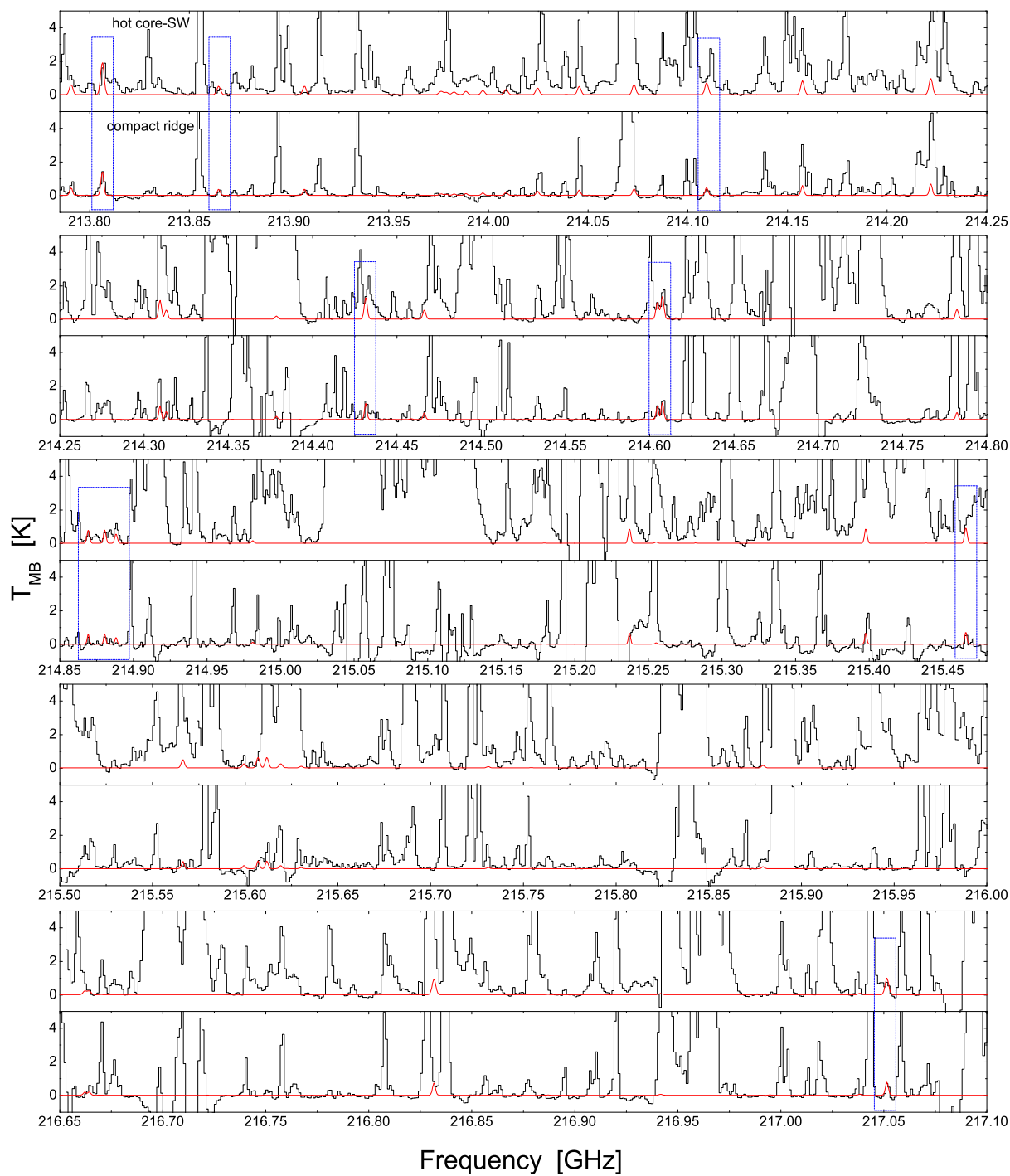}}}
\caption{Transitions of the $trans$ and $gauche$ conformers of EF (C$_{2}$H$_{5}$OCHO) from transitions 213.785 GHz to 217.100 GHz toward the hot core and compact ridge. The black curve is the observed spectra, and the red line indicates the simulated LTE spectra. Transitions with an blue box around it are listed in Table A.1.
}
\label{1}
\setcounter{figure}{\value{1}}
\end{figure*}

\clearpage
\begin{figure*}
\setcounter{6}{\value{figure}}
\setcounter{figure}{0}
\renewcommand\thefigure{A.\arabic{figure}}
\centerline{\resizebox{1.1\hsize}{!}{\includegraphics{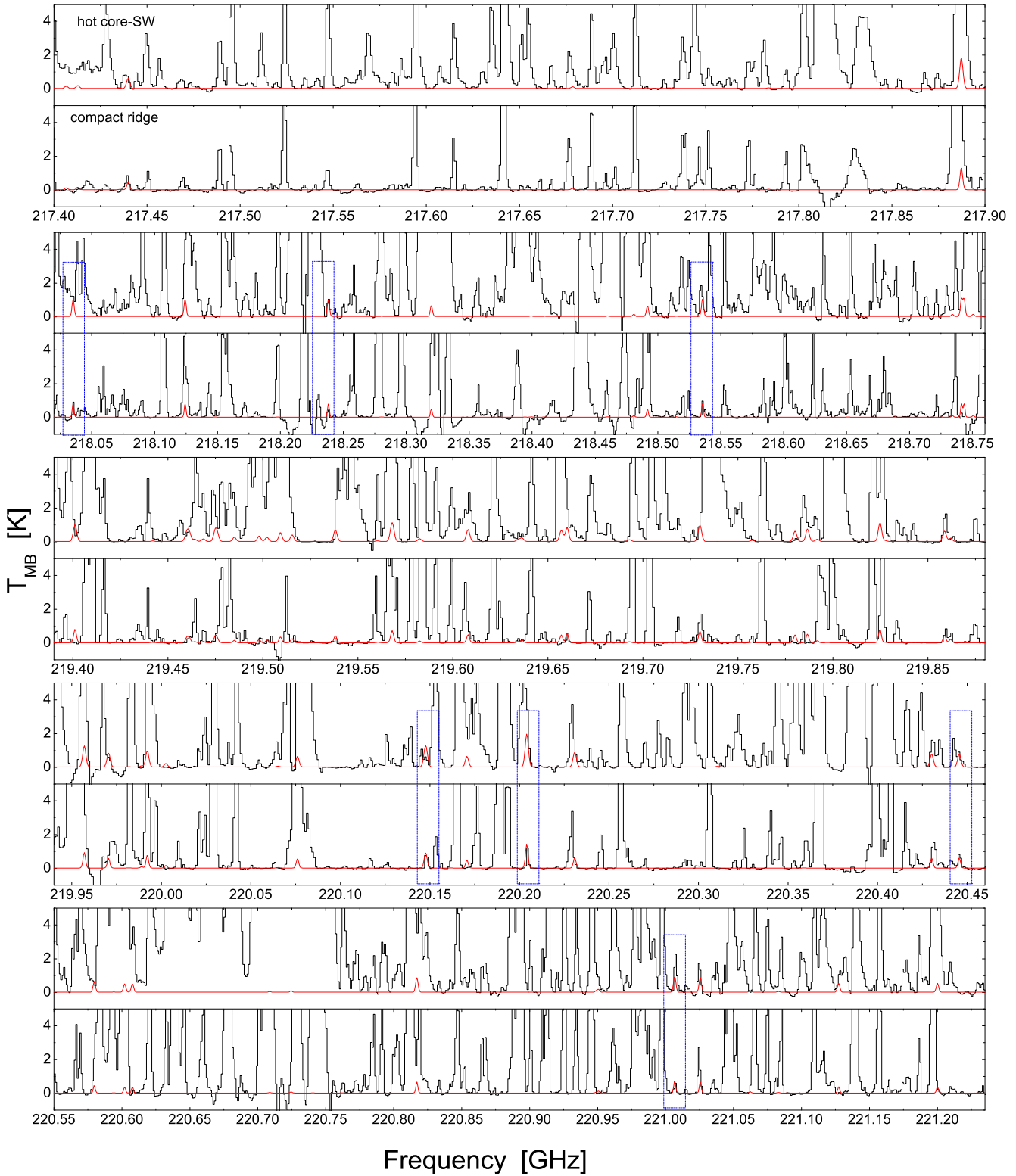}}}
\caption{Continued, from transitions 217.400 to 221.235 GHz. The spectral range of 218.76 to 219.39 Ghz is not showed in figure, because there is no EF detectable lines in this range.
}
\label{f:survey_}
\end{figure*}

\clearpage
\begin{figure*}
\setcounter{6}{\value{figure}}
\setcounter{figure}{0}
\renewcommand\thefigure{A.\arabic{figure}}
\centerline{\resizebox{1.1\hsize}{!}{\includegraphics{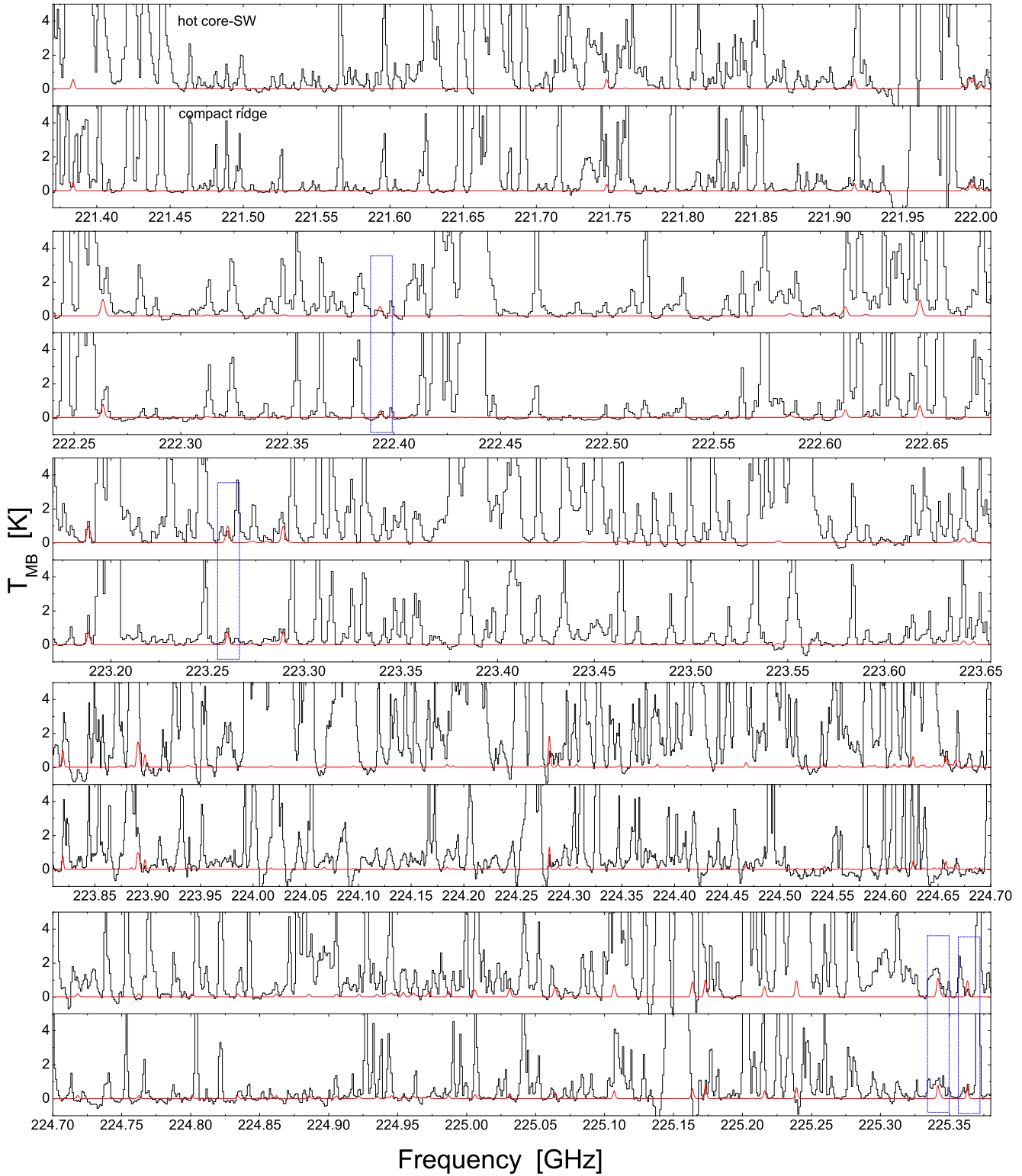}}}
\caption{Continued, from transitions 221.370 to 225.380 GHz. The range of 222.68-223.17 GHz without detectable lines is not showed.
}
\label{f:survey_}
\end{figure*}

\clearpage
\begin{figure*}
\setcounter{6}{\value{figure}}
\setcounter{figure}{0}
\renewcommand\thefigure{A.\arabic{figure}}
\centerline{\resizebox{1.1\hsize}{!}{\includegraphics{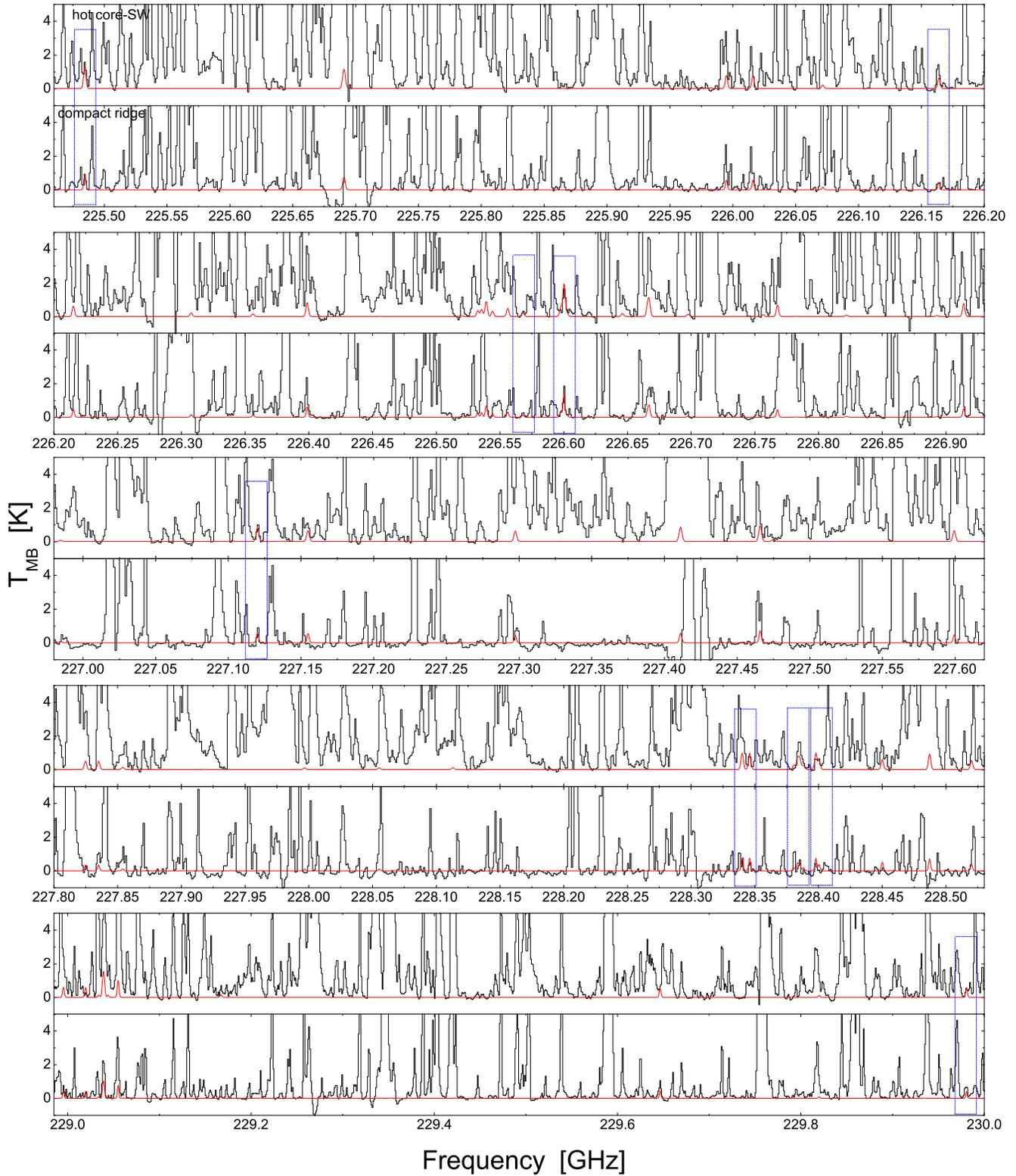}}}
\caption{Continued, from transitions 225.46 to 230.00 GHz.
}
\label{f:survey_}
\end{figure*}

\clearpage
\begin{figure*}
\setcounter{6}{\value{figure}}
\setcounter{figure}{0}
\renewcommand\thefigure{A.\arabic{figure}}
\centerline{\resizebox{1.1\hsize}{!}{\includegraphics{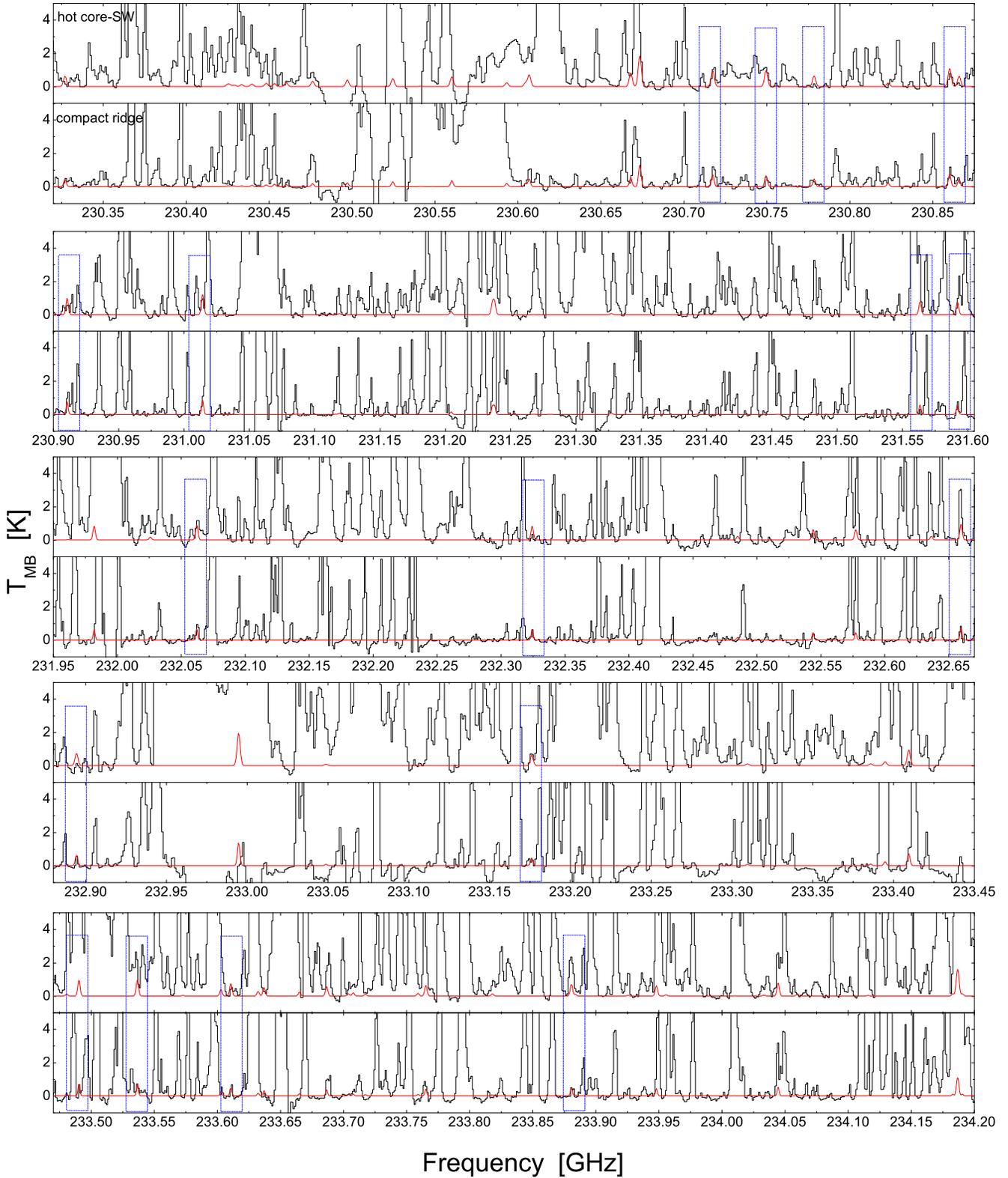}}}
\caption{Continued, from transitions 230.320 to 234.200 GHz.
}
\label{f:survey_}
\end{figure*}

\clearpage
\begin{figure*}
\setcounter{6}{\value{figure}}
\setcounter{figure}{0}
\renewcommand\thefigure{A.\arabic{figure}}
\centerline{\resizebox{1.1\hsize}{!}{\includegraphics{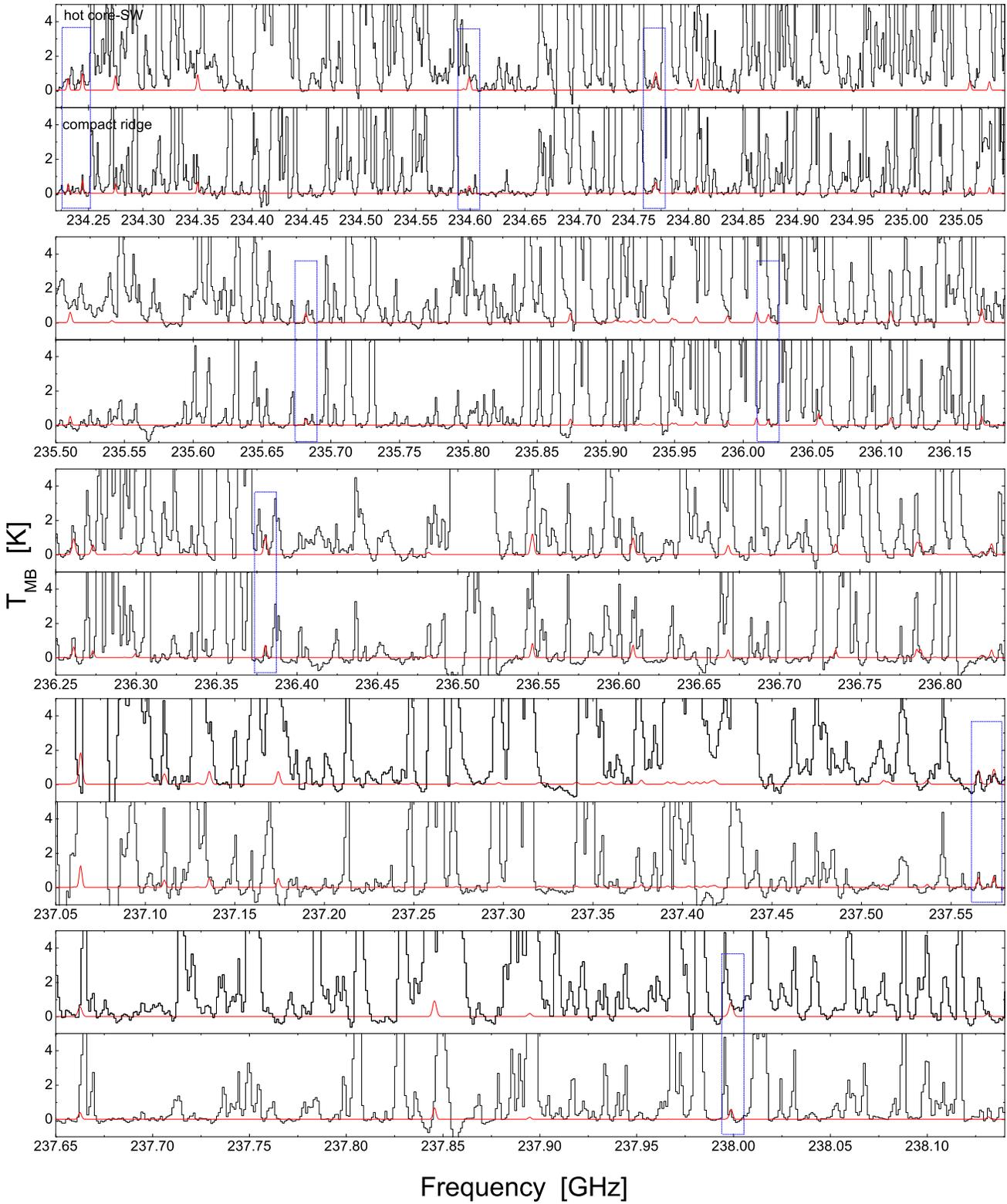}}}
\caption{Continued, from transitions 234.220 to 238.140 GHz.
}
\label{f:survey_}
\end{figure*}

\clearpage
\begin{figure*}
\setcounter{6}{\value{figure}}
\setcounter{figure}{0}
\renewcommand\thefigure{A.\arabic{figure}}
\centerline{\resizebox{1.1\hsize}{!}{\includegraphics{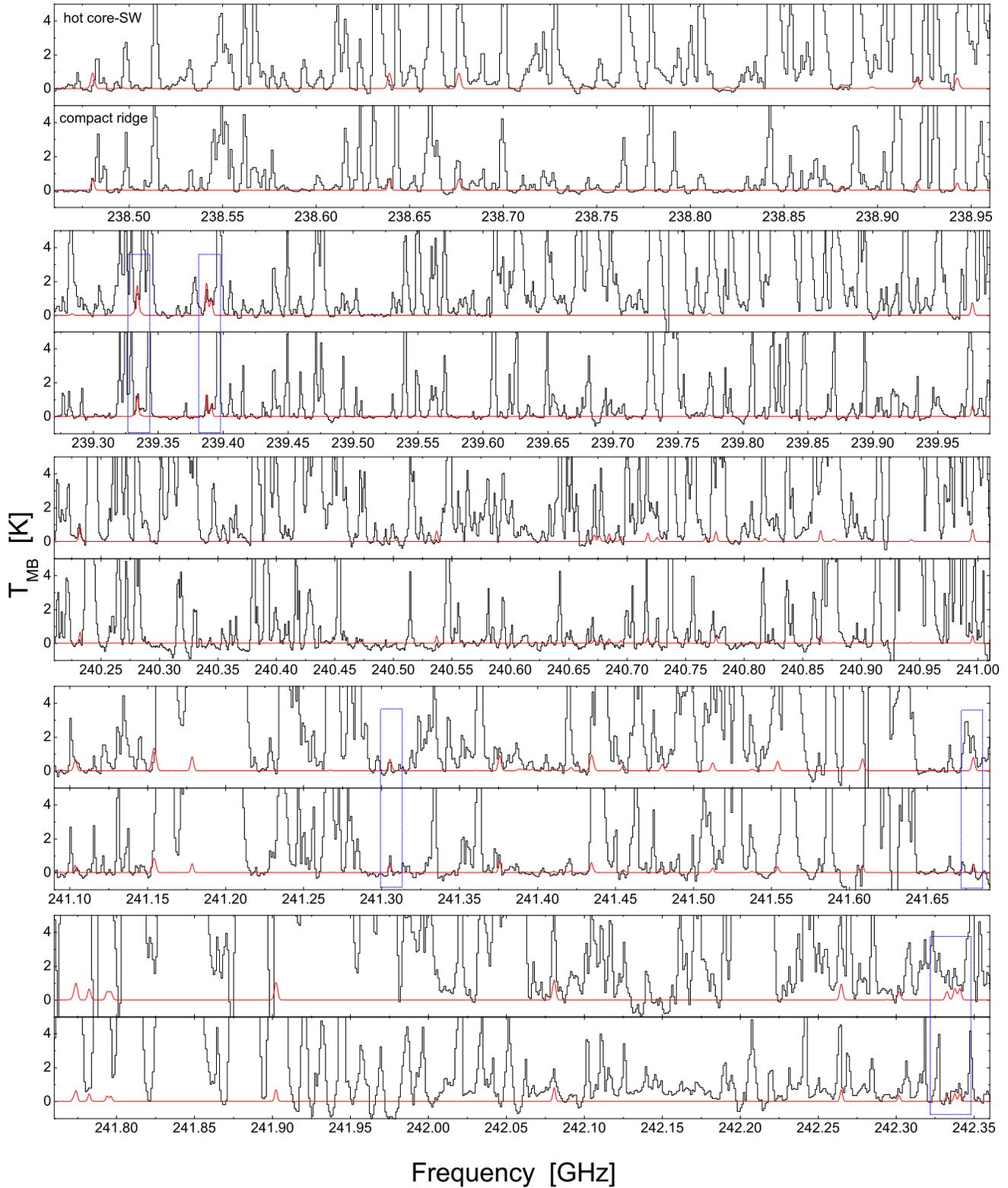}}}
\caption{Continued, from transitions 238.460 to 242.360 GHz.
}
\label{f:survey_}
\end{figure*}

\clearpage
\begin{figure*}
\setcounter{6}{\value{figure}}
\setcounter{figure}{0}
\renewcommand\thefigure{A.\arabic{figure}}
\centerline{\resizebox{1.1\hsize}{!}{\includegraphics{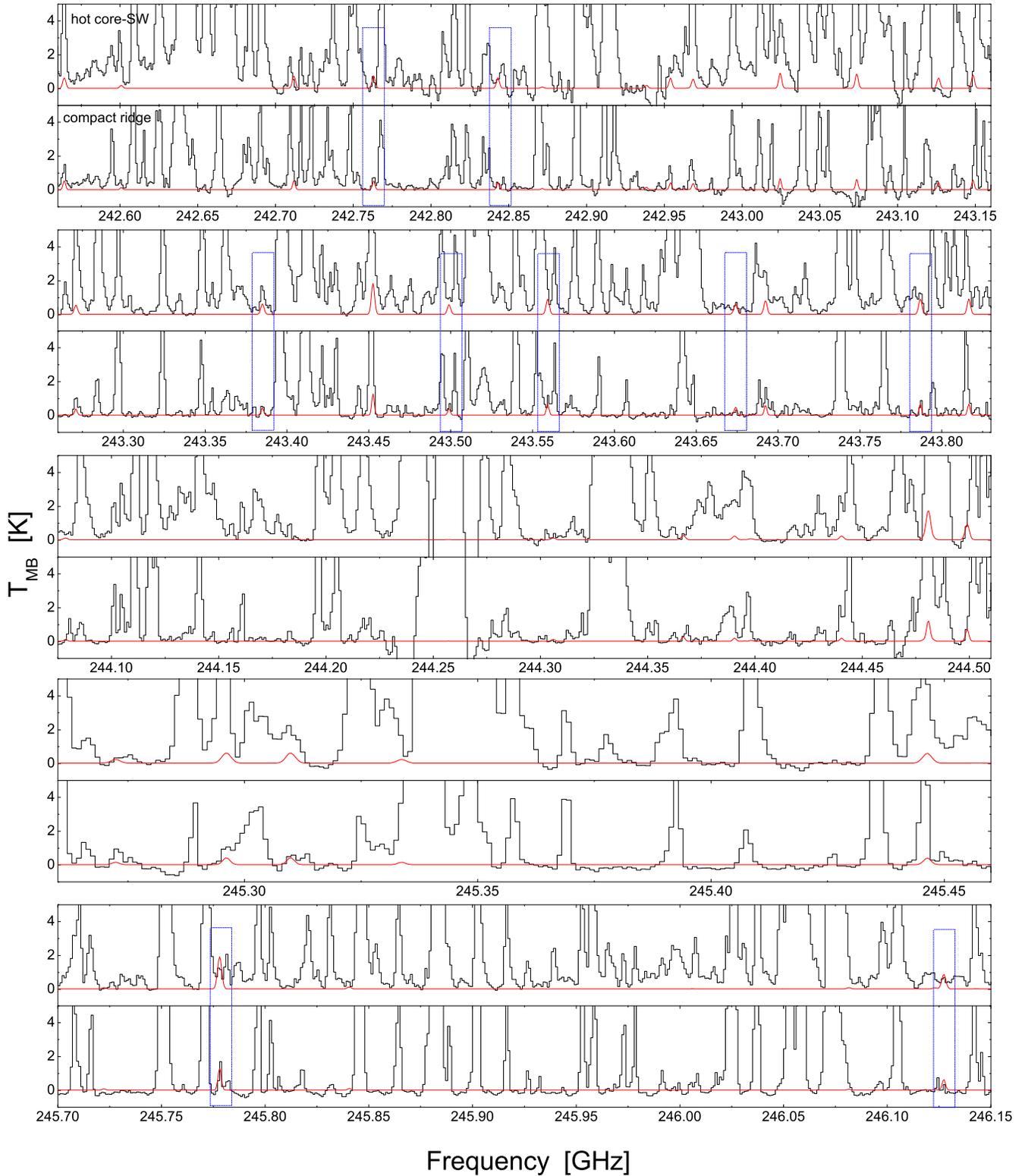}}}
\caption{Continued, from transitions 242.560 to 246.150 GHz.
}
\label{f:survey_}
\end{figure*}


\bigskip


\begin{thebibliography}{}
\bibitem[Bally et al.(2011)]{Bally11} Bally, J., Cunningham, N. J., Moeckel, N., et al. 2011, ApJ, 727, 113
\bibitem[Bally et al.(2017)]{Bally17} Bally, J., Ginsburg, A., Arce, H., et al. 2017, ApJ, 837, 60
\bibitem[Balucani et al.(2015)] {Balucani15} Balucani, N., Ceccarelli, C., \& Taquet, V. 2015, MNRAS, 449, 16
\bibitem[Belloche et al. (2009)]{Belloche09} Belloche, A., Garrod, R. T., M\"{u}ller, H. S. P., et al., 2009, A\&A, 499, 215
\bibitem[Beuther et al.(2005)]{Beuther05} Beuther, H., Zhang, Q., Greenhill, L. J., et al. 2005, ApJ, 632, 355
\bibitem[Beuther et al.(2006)]{Beuther06} Beuther, H., Zhang, Q., Reid, M. J., et al. 2006, ApJ, 636, 323
\bibitem[Blake et al.(1987)]{Blake87} Blake, G. A., Sutton, E. C., Masson, C. R., \& Phillips, T. G. 1987, ApJ, 315, 621
\bibitem[Brouillet et al.(2013)]{Brouillet13} Brouillet, N., Despois, D., Baudry, A., et al., 2013, A\&A, 550, A46
\bibitem[Brouillet et al.(2015)]{Brouillet15} Brouillet, N., Despois, D., Lu, X. H., Baudry, A., et al. 2015, A\&A, 576, A129
\bibitem[Carroll et al. (2017)]{Carroll17} Carroll, B., Crockett, N. R., Wilkins, O. H., Bergin, E. A., et al. 2017, in 72nd International Symposium on Molecular Spectroscopy
\bibitem[Cazaux et al. (2003)]{Cazaux03} Cazaux, S., Tielens, A. G. G. M., Ceccarelli, C., et al., 2003, ApJL, 593, L51
\bibitem[Chuang et al. (2016)]{Chuang16} Chuang, K. -J., Fedoseev, G., Ioppolo, S., van Dishoeck, E. F., \& Linnartz, H. 2016, MNRAS, 455, 1702
\bibitem[Crockett et al.(2014)]{Crockett14} Crockett, N. R., Bergin, E. A., Neill, J. L., et al. 2014, ApJ, 787, 112
\bibitem[Crockett et al.(2015)]{Crockett15} Crockett, N. R., Bergin, E. A., Neill, J. L., et al. 2015, ApJ, 806, 239
\bibitem[Charnley et al. (1995)]{Charnley95} Charnley, S. B., Kress, M. E., Tielens, A. G. G. M., \& Millar, T. J. 1995, ApJ, 448,232
\bibitem[Curl, R. F. (1959)] {Curl59} Curl, R. F. 1959, J. Chem. Phys., 30, 1529
\bibitem[Favre et al. (2011)]{Favre11} Favre, C., Despois, D., Brouillet, N., et al., 2011, A\&A, 532, A32
\bibitem[Favre et al. (2014)]{Favre14} Favre, C., Carvajal, M., Field, D., et al., 2014, ApJS, 215, 25
\bibitem[Favre et al. (2017)]{Favre17} Favre, C., Pagani, L., Goldsmith, P., et al., 2017, A\&A, 604, 2
\bibitem[Fedoseev et al.(2015)]{Fedoseev15} Fedoseev, G., Ioppolo, S., Zhao, D., Lamberts, T., \& Linnartz, H. 2015, MNRAS, 446, 439
\bibitem[Feng et al.(2015)]{Feng15} Feng, S., Beuther, H., Henning, Th., et al. 2015, A\&A, 581, 71
\bibitem[Friedel \& Snyder(2008)]{Friedel08} Friedel, D. N., \& Snyder, L. E. 2008, ApJ, 672, 962
\bibitem[Friedel \& Looney(2017)]{Friedel17} Friedel, D. N., \& Looney, L. W. 2017, AJ, 154, 152
\bibitem[Fontani et al. (2007)]{Fontani07} Fontani, F., Pascucci, I., Caselli. P., Wyrowski, F., Cesaroni, R., Walmsley, C. M., 2007, A\&A, 470, 639
\bibitem[Garrod \& Herbst (2006)]{Garrod06} Garrod, R. T., \& Herbst, E. 2006, A\&A, 457, 927
\bibitem[Garrod et al. (2008)]{Garrod08} Garrod, R. T., Weaver, S. L. W., Herbst, E., 2008, ApJ, 682, 283
\bibitem[Greenhill et al.(2004)] {Greenhill04} Greenhill, L. J., Gezari, D. Y., Danchi, W. C., et al. 2004, ApJ, 605, L57
\bibitem[Goddi et al.(2011)]{Goddi11} Goddi, C., Greenhill, L. J., Humphreys, E. M. L., Chandler, C. J., \& Matthews, L. D. 2011, ApJL, 739, L13
\bibitem[Gong et al.(2015)]{Gong15} Gong, Y., Henkel, C., Thorwirth, S., et al., 2015, A\&A, 581, 48
\bibitem[G\'{o}mez et al.(2008)]{Gomez08} G\'{o}mez, L., Godr\'{i}guze, L. F., Loinard, L., et al. 2008, ApJ, 685, 333
\bibitem[Gu\'{e}lin et al.(2008)]{Guelin08} Gu\'{e}lin, M., Brouillet, N., Cernicharo, J., Combes, F., \& Wootten, A. 2008, Ap\&SS, 313, 45
\bibitem[Hirota et al. (2017)]{Hirota17} Hirota, T., Machida, M. N., Matsushita, Y., et al. 2017, Nature Astronomy, 1, 0146
\bibitem[Lis et al. (1991)]{Lis91} Lis, D. C., Carlstrom, J. E., \& Keene, J. 1991, ApJ, 380, 429
\bibitem[Liu et al. (2002)]{Liu02} Liu, S.-Y., Girart, J. M., Remijan, A., Snyder, L. E. 2002, ApJ, 576, 255
\bibitem[Marcelino et al. (2018)]{Marcelino18} Marcelino, N., Gerin, M., Cernicharo, J., Fuente, A., et al. 2018, A\&A, 2018arXiv180908014M
\bibitem[Medvedev et al. (2009)]{Medvedev09} Medvedev, I. R., De Lucia, F. C., Herbst, E. 2009, ApJS, 181, 433
\bibitem[Menten \& Reid.(1995)]{Menten95} Menten K. M., \& Reid, M. J. 1995, ApJ, 445, 157
\bibitem[Millar et al. (1991)]{Millar91} Millar, T. J., Bennett, A., Rawlings, J. M. C., et al., 1991, A\&AS, 87, 585
\bibitem[M\"{o}ller et al.(2013)]{Moller13} M\"{o}ller, T., Bernst, I., Panoglou, D., et al. 2013, A\&A, 549, 21
\bibitem[M\"{o}ller et al.(2017)]{Moller17} M\"{o}ller, T., Endres, C., Schilke, P. 2017, A\&A, 598, 7
\bibitem[M\"{u}ller et al. (2001)]{Muller01} M\"{u}ller, H. S. P., Thorwirth, S., Roth, D. A.,\& Winnewisser, G. 2001, A\&A, 370, 49
\bibitem[M\"{u}ller et al. (2005)]{Muller05} M\"{u}ller, H. S. P., Schl\"{o}der, F., Stutzki, J., \& Winnewisser, G. 2005, Journal of Molecular Structure, 742, 215
\bibitem[Pagani et al. (2017)]{Pagani17} Pagani, L., Favre, C., Goldsmith, P. F., et al., 2017, A\&A, 604, 32
\bibitem[Peng et al. (2017)]{Peng17} Peng, Y. P., Qin, S. L., Schilke, P., et al., 2017, ApJ, 837, 49
\bibitem[Peng et al.(2013)]{Peng13} Peng, T.-C., Despois, D., Brouillet, N., et al. 2013, A\&A, 554, A78
\bibitem[Pickett et al. (1998)]{Pickett98} Pickett, H. M., Poynter, R. L., Cohen, E. A., et al. 1998, Quant. Spectrosc. \& Rad. Transfer, 60, 883
\bibitem[Qin et al. (2010)]{Qin10} Qin, S.-L., Wu, Y., Huang, M., et al. 2010, ApJ, 711, 399
\bibitem[Riveros \& Wilson. (1967)]{Riveros67} Riveros, J. M., \& Wilson, E. B. 1967, JChPh, 46, 4605
\bibitem[Rivilla et al. (2013)]{Rivilla13} Rivilla, V. M., Mart\'{i}n-Pintado, J., Jim\'{e}nez-Serra, I., Rodr\'{i}guez-Franco, A. 2013, A\&A, 554, A48
\bibitem[Rivilla et al. (2017a)]{Rivilla17a} Rivilla, V. M., Beltr\'{a}n, M. T., Cesaroni, R., et al., 2017a, A\&A, 598, 59
\bibitem[Rivilla et al. (2017b)]{Rivilla17b} Rivilla, V. M., Beltr\'{a}n, M. T., Mart\'{i}n-Pintado, J., et al., 2017b, A\&A, 599, 26
\bibitem[Rizzo et al. (2017)]{Rizzo17} Rizzo, J. R., Tercero, B., \& Cernicharo, J. 2017, A\&A, 605, A76
\bibitem[Sakai et al. (2015)]{Sakai15} Sakai, Y., Kobayashi, K., \& Hirota, T. 2015, ApJ, 803, 97
\bibitem[Sault et al.(1995)]{Sault95} Sault, R. J., Teuben, P. J., \& Wright, M. C. H. 1995, in ASP Conf. Ser. 77, Astronomical Data Analysis Software and Systems IV (San Francisco: ASP), 433
\bibitem[Shuping et al.(2004)]{Shuping04} Shuping, R., Morris, M., \& Bally, J. 2004, AJ, 128, 363
\bibitem[Simpson et al.(2006)]{Simpson06} Simpson, J. P., Colgan, S. W. J., Erickson, E. F., et al. 2006, ApJ, 642, 339
\bibitem[Snyder \& Lovas.(2005)]{Snyder05}Snyder, L. E., Lovas, F. J., Hollis, J. M., et al. 2005, ApJ, 619, 914
\bibitem[Taquet et al. (2016)]{Taquet16} Taquet, V., Wirstr\"{o}m, E. S., Charnley, S. B. 2016, ApJ, 821, 46
\bibitem[Tercero et al. (2015)]{Tercero15} Tercero, B., et al., 2015, A\&A, 582, L1
\bibitem[Tercero et al. (2013)]{Tercero13} Tercero, B., Kleiner, I., Cernicharo, J., et al., 2013, ApJL, 770, L13
\bibitem[Tercero et al.(2010)] {Tercero10}Tercero, B., Cernicharo, J., Pardo, J. R., \& Goicoechea, J. R. 2010, A\&A, 517, A96
\bibitem[Wang et al.(2011)]{Wang11} Wang, S., Bergin, E. A., Crockett, N. R., et al. 2011, A\&A, 527, A95
\bibitem[Widicus Weaver \& Friedel(2012)]{Widicus Weaver12} Widicus Weaver, S. L., \& Friedel , D. N. 2012, ApJS, 201, 16
\bibitem[Wright \& Plambeck(2017)]{Wright17} Wright, M. C. H., \& Plambeck, R. L. 2017, ApJ, 843, 83
\bibitem[Wu et al.(2014)]{Wu14} Wu,Y.F., Liu, T., \& Qin, S. L. 2014, ApJ, 791, 123
\bibitem[Zapata et al.(2011)]{Zapata11} Zapata, L. A., Schid-Burgk, J., \& Menten, K. M. 2011, A\&A, 529, A24
\bibitem[Zapata et al.(2012)]{Zapata12} Zapata, L. A., Rodr\'{i}guez, L. S, Schmid-Burgk, J., et al. 2012, ApJL, 754, 17
\bibitem[Zhang et al.(2014)]{Zhang14} Zhang, C.-P., Wang, J.-J., Xu, J.-L., et al., 2014, ApJ, 784, 107

\end{thebibliography}
\end{document}